\documentstyle[aps,epsf,eqsecnum,pre]{revtex}

\begin{document}

\def \be #1{\begin{equation}\label{#1}}
\def \ee {\end{equation}}
\def \bea #1{\begin{eqnarray}\label{#1}}
\def \eea {\end{eqnarray}}
\def \Eq #1{Eq.~(\ref{#1})}
\def \Fig #1{Fig.~\ref{#1}}
\def \q {{\bf q}}

\title{Nonlinear Measures for Characterizing Rough Surface Morphologies}

\author{Jan\'{e}  Kondev\cite{new_addJK}}
\address{Institute for Advanced Study, Olden Lane, Princeton, NJ 08540, and \\
      Department of Physics, Princeton University, Princeton, NJ 08540}

\author{Christopher L.  Henley and David G. Salinas\cite{new_addDS}}
\address{Laboratory of Atomic and Solid State Physics, Cornell University,
Ithaca, New York, 14853}

\date{\today}

\maketitle

\begin{abstract}
We develop a  new approach to characterizing the morphology of rough 
surfaces based on the analysis of the scaling properties of contour 
loops, i.e. loops of constant height. Given a height profile of the surface 
we perform independent measurements of the fractal dimension of 
contour loops, and the exponent that characterizes their size distribution.
Scaling formulas are derived and used to relate these two    
geometrical exponents  to the roughness exponent of a self-affine surface, 
thus providing independent measurements of this important quantity.  
Furthermore, we define the scale dependent curvature and demonstrate that 
by measuring its third moment departures of the height fluctuations 
from Gaussian behavior can be ascertained.   
These nonlinear measures are used to characterize the morphology 
of computer generated Gaussian rough surfaces, surfaces obtained in 
numerical simulations of a simple growth model, and surfaces observed  
by scanning-tunneling-microscopes.   
For experimentally realized surfaces  the self-affine scaling 
is cut off by a correlation length, and we generalize our theory of 
contour loops to take this into  account. 
\end{abstract}


\section{Introduction}

Random surfaces are widely used in the physical sciences
to model  phenomena ranging from the extremely 
small (quantum gravity) to  the very large  (Earth's relief). 
They describe crack fronts in materials science \cite{bouchaud},
ripple-wave turbulence \cite{putterman}, 
passive tracers in two-dimensional fluid flows \cite{gollub,tabeling},  
cloud perimeters \cite{lovejoy,jon}, shapes of stromatolites
(conjectured fossil accretions of ancient bacteria)\cite{stomatolites}, 
to mention but a few recent examples.  
Our focus in this paper is on the morphology of deposited metal films,
which develop random self-affine surfaces
under several quite different non-equilibrium growth conditions, 
as indicated by theoretical, numerical, and experimental results 
over the past decade\cite{krimrev,krugrev}. 

Surface configurations are parametrized by a two-dimensional 
field $h({\bf x})$ which represents  the height of the surface above a
reference plane $\{\bf{x}\}$.
Theoretically the  dynamics of a growing surface are  
described by a continuum (Langevin) 
equation giving $d h({\bf x})/dt$ as a sum of 
a Gaussian white noise term, to mimic the random deposition of atoms, 
and a polynomial of various gradients of $h({\bf x})$, to 
model relaxation processes on a coarse grained scale. 
The nonequilibrium growth behavior is due to the  interplay of 
the deposition and relaxation terms.

Different relaxation terms are appropriate for different growth 
conditions, and the notion of universality has been taken over 
from the study of critical phenomena. Namely, it is believed that 
there are only a few distinct universality classes of growth 
characterized by exponents which describe the temporal and 
spatial scaling of the growing interface. 

One would guess that even a snapshot of the morphology
should carry the evidence of the non-equilibrium, nonlinear 
growth process which produced it, 
and should differ measurably from surfaces produced 
in equilibrium or by a linear process, even if
they share the same scaling exponents. 

This motivates a search for roughness measures independent
of the quadratic ones (e.g. the rms height), 
which might identify important distinctions
between different surface models that have similar spatial power spectra. 
Such measures, although motivated in the context of self-affine
or multi-affine surfaces, should be handy even for surfaces 
showing no self-affine regime. 
They can quantify features of morphology which are presently 
characterized by eye, which should permit a more systematic 
comparison between observations and models than at present. 
One can imagine that, armed with two or three kinds of roughness measures
tuned to different qualitative aspects of the surfaces morphology, one could
construct empirical ``phase diagrams'' in this two or three-dimensional 
parameter space, e.g. mapping out domains in the parameter space that
correspond to various growth conditions. 

Given the surface as parametrized by an array of heights --
obtained, e.g., from a simulation or a scanning-tunneling-microscope (STM)
experiment \cite{gibson} -- we ask, in what
different ways can the surface morphology be characterized?  
In general, one requires more than one characterization to confirm a 
match between experimental and simulation data, or to convincingly
verify self-affineness.
For the applied problems of growing flat surfaces (e.g. for semiconductor
devices) or regularly modulated ones (e.g. to nanofabricate arrays
of quantum dots), it is also desirable to develop independent measures 
that quantify different aspects of a rough surface's geometry.
In this paper, we propose two categories of novel measure 
for characterizing spatial correlations of rough surfaces. 
These measures are usable on any kind of
rough model, and require no dynamical information, so they
should be useful in analyzing not only experiments on
solid films, but all the diverse phenomena mentioned above.

\subsection{Outline of the paper}

We start with a short review (Section~\ref{sec_affine})
of self-affine geometry in terms of its
real-space, Fourier space and fractal properties. 
In section \ref{sec_measures}
we introduce the non-linear measures,  the scale dependent 
curvature and the loop  measures, for which 
various scaling relations are derived in section \ref{sec_scaling}. These
scaling relations are modified in the presence of a (time dependent) cutoff 
length-scale above 
which the height fluctuations are no longer self-affine, and this situation 
is described towards the end of section \ref{sec_scaling}. This concludes 
the first half of the paper which deals with 
the theory of non-linear measures. 

The second half of the paper is devoted to analyzing data obtained in
numerical simulations and experiments {\em using} the measures introduced
in the first part. It starts off with 
section \ref{sec_gauss} where we present the results of our simulations of
random Gaussian surfaces for various values of the roughness exponent 
$0\le\alpha\le1$. 
These simulations serve to confirm the various 
scaling relations derived earlier. 
In section \ref{sec_SSM} the non-linear measures 
are applied to a non-equilibrium 
growth model, the so-called single step model, 
which is known to belong to the
KPZ universality class. Finally, in section \ref{sec_exp} 
we demonstrate the usefulness of our measures for analyzing experimental 
data on the example provided by an STM image of a growth 
roughened metal film. The discussion section (section \ref{sec_disc}) 
summarizes our main results, gives a critical comparison between the newly 
introduces measures
and those used previously, and points out some interesting new directions in 
which progress can be made. The three appendices are reserved for details
of the calculation of the loop correlation exponent in the case of 
equilibrium rough surfaces (appendix~\ref{app_O2loop}), 
details of the derivation of percolation 
exponents for contours  of uncorrelated heights (appendix~\ref{app_perc}), 
as well as a full description of the loop finding algorithm which is at the 
heart of the numerical  simulations and the loop analysis of 
STM data (appendix~\ref{app_loopalg}).

\section{Self-affine geometry} 
\label{sec_affine}

Here we review the scaling properties of self-affine interfaces
in real-space and in  Fourier-space, as well as  the
fractal geometry of their level sets.
The surface is fully specified by the height
field $h({\bf x})$, which 
may be the microscopic heights of individual
surface atoms above the substrate, as measured by an STM, 
or it may be a 
coarse grained quantity representing the average of  individual atomic 
heights over a region \cite{FN-nooverhang}.

The defining property of {\it self-affine} surfaces is their invariance 
under rescaling. Namely, 
the probability distribution function (PDF) for $h({\bf x})$ is such that 
\be{affinedef}
h({\bf x}) \cong b^{-\alpha} h(b {\bf x}) 
\ee
for any $b>1$, where  $\alpha$ 
is the {\em roughness exponent}; here 
the symbol $\cong$  means 
``statistically equivalent with respect to the PDF''. 
In other words, if we stretch the surface by a rescale factor  $b$ 
in the horizontal direction 
(= parallel to the reference plane ${\bf x}$), 
then to obtain a statistically equivalent surface,
we must stretch by a factor $b^\alpha$ in the vertical direction 
(= the perpendicular direction of the heights $h({\bf x})$). 
A central theme of this paper is 
the different ways of determining $\alpha$, given a height 
profile $h({\bf x})$. 

A self-affine surface is {\em rough} if $\alpha>0$. 
Furthermore, for the surface  to exhibit  a two-dimensional 
character (at distances much larger than the surface width)  
the rescale factor in the vertical direction ($b^\alpha$) can not
exceed the one in the horizontal direction ($b$), i.e., 
we require $\alpha \le 1$.

\subsection{Real-space properties}

The self-affine scaling of the height is typically measured by  the 
height-correlation function 
\be{struc} 
D_2({\bf r}) =   \langle [h({\bf x}+{\bf r}) - h({\bf x})]^2\rangle 
  \sim |{\bf r}|^{2 \alpha} \ , 
\ee
where the scaling with separation $|{\bf r}|$ is  a direct 
consequence of the self-affine property, \Eq{affinedef}, which states that 
$h$ has a scaling dimension $\alpha$. 

In experiments the correlations that lead to self-affine scaling of the 
surface develop over time, which we take to be  measured from the start 
of the deposition  process. Namely, 
after time $t$ self-affine scaling will be observed only  up-to 
length scales smaller then the {\em correlation length} $\xi(t)$. Physically 
the height correlations develop due to the various surface relaxation 
processes that are present under the given growth conditions.
 
Numerical simulations of various surface growth models,
as well as experiments under different conditions, have 
shown that $\xi$  grows with the duration of the deposition process, $t$,  
according to the dynamical scaling relation \cite{barabasi}:
\be{dynamical}
\xi (t) \sim  t^{1/z} \; .
\ee 
It is believed that there are
only a few different universality classes of growth each characterized 
by the exponents $\alpha$ and $z$ \cite{krugrev,barabasi}.
Experimental efforts have 
been focused on extracting these exponents from data obtained using
various surface-sensitive methods: X-rays or helium diffraction,
STM scans, etc.\cite{krimrev} 
In this paper, we will be almost entirely concerned with 
the spatial (equal-time) correlations.

\subsection{Fourier-space properties}

The power spectrum of a self-affine surface 
\be{power}
S({\bf q}) = \langle |\tilde{h}({\bf q})|^2\rangle 
\ee
is defined in terms of the Fourier transformed height
\be{FTg}
\tilde{h}({\bf q}) = \int \! d^2 \! {\bf x} \: h({\bf x}) 
                    e^{-{\rm i}{\bf q}\cdot{\bf x}} \;. 
\ee

The height correlation function is {\em linearly} 
related to the height power spectrum, 
\be{struc2}
 D_2({\bf r}) = \int \! d^2 \! {\bf q} \: S({\bf q})
	   (e^{{\rm i}{\bf q}\cdot{\bf r}} - 1) \; ,     
\ee
as is any other translation-invariant expectation 
quadratic in heights, such as the net variance.  
\Eq{struc2} and \Eq{struc} imply the scaling
\be{hqvar}
   S({\bf q}) \sim |{\bf q}|^{-2 (1+\alpha)} \; , 
\ee
for small values of $|{\bf q}|$.
In the case of a surface with a finite correlation length, 
$S({\bf q})$ crosses over to a constant value for
$|{\bf q}|<1/\xi(t)$; this situation is discussed in more detail
in Sec.~\ref{subsec_perc}.

Clearly, $S({\bf q})$ does not uniquely characterize a self-affine
ensemble of surfaces. 
For example, it is invariant under $h({\bf x}) \to - h({\bf x})$, 
yet surfaces produced in non-equilibrium growth typically
break the up/down symmetry. 
Furthermore, given any $S({\bf q})$ one can always  construct
a Gaussian ensemble by linear addition of Fourier components
-- we do this in Sec.~\ref{sec_gauss} -- yet the real growth 
process is typically nonlinear, and the surface is non-Gaussian.
Indeed, confirmation of the scaling given by Eqs.~\ref{struc}
or~\ref{hqvar} in experiments can {\em not} be 
interpreted as conclusive evidence for a self-affine geometry: 
that is a property of the whole ensemble 
and so requires proper scaling of all moments and correlations,
not just  the second moment.

\subsubsection {Quadratic roughness measures}
\label{subsec_quadmeas}

The quantities (\ref{struc}) and (\ref{power})
are quadratic measures of roughness, which we shall also 
call ``linear''; untill recently, no other
kind was in use.

The height-correlation function, \Eq{struc}, 
is the most standard measure in theoretical discussions, in that
``roughness'' is defined by the divergence of this function as
its argument $\bf r$ approaches infinity. 
(Non-monotonic behavior of this function has also been used \cite{cahill}
to measure the characteristic spatial scale of mounds or other
patterns in {\it non}-self-affine surfaces.)

On the other hand, the Fourier power spectrum (\ref{power})
is central in theoretical derivations but rarely used in experimental
analysis (except for Ref.~\onlinecite{williams}). 
This seems to be the best quadratic measure,  in that it most cleanly 
separates the contributions from fluctuations on different length scales, 
and it shows the sharpest knee (in a log-log plot) where self-affine
scaling is cut off. 

Another quadratic measure is  the total variance 
of $h({\bf x})$ in a in a box of size $b$, 
as a function of $b$ \cite{fourier_patch,schmitt}:
\be{roughness-var}
  \langle (h({\bf x})-\overline{h}_b)^2\rangle _b, 
\end{equation}
where $\overline{h}_b\equiv
\langle h({\bf x})\rangle_b$, and $\langle \ldots \rangle _b$ means
the spatial 
average is only taken over a square of side $b$ centered on ${\bf x}_0$;
this variance should be averaged over different choices of ${\bf x}_0$.

\subsection{Fractal properties}

Self-affine surfaces are fractals only in a generalized sense,
since the horizontal direction 
rescales differently from the vertical direction.
On the other hand, the level set of such a surface 
(defined as its intersection with a horizontal plane)
{\it is} a fractal object~\cite{mandelbrot}; see \Fig{fig_loops}, below.
Different planes of intersection give statistically
equivalent level sets, since the height fluctuations of a rough 
surface are unbounded. Level sets consist of contour loops which 
are the connected components. We expect these to be fractal as well, 
with a fractal dimension {\em smaller} then the dimension 
of the whole level set, which is simply  the union of all  
contour loops of the same height. 
Furthermore, contour loops come in all sizes limited only by the 
system size, and an exponent 
can be defined that characterizes their size distribution. Since contour 
loops are connected clusters their 
{\em geometrical} exponents are analogous to those defined for  
percolation clusters. 

We will show that 
the scaling of contour loops uniquely specifies the scaling of the
associated self-affine rough surface; this will be expressed in formulas
giving the geometrical exponents in terms of the roughness exponent $\alpha$.
It is somewhat surprising that, 
by doing measurements solely on the level set, information can be 
obtained about the out-of-plane fluctuations of the surface. 
For experiments that yield
only level-set data without the heights, 
(e.g. freeze-fracture electron microscopy \cite{freeze})     
our contour-loop analysis is the {\it only} route to
extracting the roughness exponent.

\section{Nonlinear measures}
\label{sec_measures}

In the past the analysis of rough surfaces mostly relied on 
measures which probed the  second moment of the heights,  such as  the 
height-correlation function or the power spectrum
(defined in \Eq{struc} and \Eq{hqvar} below).
But that is inherently insufficient to distinguish different
growth ensembles or even to verify self-affineness.

Therefore, to more fully characterize rough surfaces we introduce
in this section
two new types of {\em non-linear} measures, i.~e., measures that are 
{\em not} linearly related to the structure function of the height field.  
Non-linear measures of the first type (Sec.~\ref{subsec_curv}) are 
moments of the ``scale-dependent curvature'', 
a modification of the standard height correlation function which
can identify deviations from Gaussianness of the height fluctuations, 
in particular the skew (up/down asymmetry) at various length scales. 
(We will compare these to existing non-quadratic roughness
measures in the Discussion part of  Sec.~\ref{sec_disc}.)

Measures of the second type were introduced in 
Ref.~\onlinecite{loop-PRL}; they
are distributions of three different geometrical quantities 
defined for  contour-loops (or simply ``loops'') 
of constant height, which make up the level sets of the height function. 
These measures are associated with geometrical exponents that 
characterize contour loops on self-affine rough surfaces: the 
loop correlation exponent, the fractal dimension of a loop, and the 
length distribution exponent.

\subsection{Scale-dependent curvature} 
\label{subsec_curv}

The obvious real-space-based nonquadratic 
generalization of the height-correlation function
is 
  \be{cubicdiff}
   \langle [(h({\bf x}+{\bf r})-h({\bf x})]^3 \rangle ;
   \ee
however, this is identically zero on an isotropic surface
(and whenever ${\bf r} \to -{\bf r}$ is a symmetry).
To escape this problem, we observe that $h({\bf x}+{\bf r})-h({\bf x})$
is a sort of first difference at scale $r$, and 
replace it by a sort of second difference.
Namely, we define the ``curvature at $\bf x$ on scale $b$'' as
\be{curv_def}
C_b({\bf x}) = \sum _{m=1}^M [h({\bf x}+b{{\bf e}_m})-  h({\bf x}) ]
\ee
where the offset directions 
$\{{\bf e}_m \}$ are a fixed set of vectors summing to zero.
In our numerical implementation of this measure, 
where $\{ \bf x \}$ is a square lattice,  
we choose four such offsets related by $90^\circ$ rotations, 
pointing either along the $\{10\}$ or the $\{11\}$ type directions.
Those two sets of offsets should give equivalent results (for the same $b$),
provided the surface is statistically invariant under rotations in the
reference plane.  We then define curvature moments $\langle C_b^q \rangle$ 
for integer powers $q$. 

The first moment of $C_b$ is manifestly zero;
the second moment is linearly related to the height-correlation
function: 
\be{curv2}
   \langle C_b({\bf x})^2 \rangle  = 
   M \sum _{m=1}^{M} D_2(b {\bf e}_m ) - \frac{1}{2} \sum _{m,n=1}^{M} 
   D_2(b({\bf e}_m-{\bf e}_n)) \ .
\ee
(This is shown by inserting \Eq{curv_def} and then decoupling each term
of the double sum using the identity $(h_m-h_0)(h_n-h_0)= 
{1 \over 2} \{ (h_m-h_0)^2 + (h_n-h_0)^2 - (h_m-h_n)^2 \}$.)

The higher moments of $C_b$ serve to measure the (possible)
deviation of the height fluctuations from the Gaussian distribution. 
For example, if the surface has up/down symmetry $h \leftrightarrow -h$  
(as all Gaussian surfaces do), $\langle[C_b({\bf x})]^3\rangle$  vanishes.
On the other hand, non-equilibrium grown surfaces often have 
rounded ``hilltops'' and sharp ``valleys'';
that tends to make $\langle C_b^3 \rangle > 0$, a signature of ``skew''
in the distribution.
Similarly, the fourth moment can also be used to test
whether the surface is Gaussian, since in that case
\be{kurt}
{\langle[C_b({\bf x})]^4\rangle}/ {\langle[C_b({\bf x})]^2\rangle^2} 
= 3 \ .
\ee

For a self-affine surface
\be{curv_scale}
\langle[C_b({\bf x})]^q\rangle \simeq {\rm const} |{\bf x}|^{q \alpha} 
	\ee
follows from \Eq{affinedef};
of course the coefficient might be zero as is the case 
for odd $q$, when the height field has up/down symmetry. 

Functions such as $\langle C_b^q  \rangle$ (as a function of $b$)
or $D_q(r)$ (as a function of $r$) can also be used as ``spectra''
of the height fluctuations for {\it non} self-affine surfaces.
That is, 
differences in the  behavior of the function in different ranges of $b$ or $r$ 
reveal qualitative differences of the surface morphology on the
corresponding length scales. 

In principle,  $q$-th order moments 
may scale with well-defined exponents $\alpha_q$, yet the surface is
not self-affine since $\alpha_q\neq q \alpha$ violating \Eq{curv_scale};
this is called a ``multifractal'' or, more precisely, ``multiaffine''
surface \cite{krug_PRL}.
Slow transients  of multiaffine behavior
(up to $\sim 10^8$ steps in $d=1+1$ and $\sim 10^3$ steps in $d=2+1$)
have been seen recently in numerical simulations of growth
models \cite{DS-punyindu}
(which, however, are believed to be asymptotically self-affine). 
The analogous higher order structure functions 
are a central issue in turbulence, 
where the violation of self-affine 
(Kolmogorov) scaling is well established and is 
associated with intermittency of the velocity 
field fluctuations\cite{sreeni}.

The scale dependent curvature 
can be contrasted with Krug's height-difference moments\cite{krug_PRL}, 
\be{Krugmoment}
   D_q({\bf r}) \equiv \langle |h({\bf x}+{\bf r})-h({\bf x})|^q\rangle, 
\ee
a natural generalization of the height-correlation function 
using an absolute value to avoid the trivial cancellation  in 
\Eq{cubicdiff}. 
Das Sarma and collaborators\cite{dassarma1,DS-punyindu} used
(\ref{Krugmoment}) to
test for multi-affine
behavior (whereby the $1/q$ power of the $q$ moment scales with exponent
{\it depending on} $q$, unlike the simpler self-affine case).
For odd $q$, \Eq{Krugmoment} is insensitive to the
up-down symmetry (or lack thereof) since it is nonzero anyhow.
Our ``curvature'' seems to be the simplest function that
detects the skew locally.

\subsection {Fractal dimension of contour loops}
\label{scaling-dimension}

For the remainder of this section, we must define the {\it loop
ensemble}.
Consider a contour plot of a rough surface with a fixed
spacing $\Delta$ between heights of successive level sets. 
We take it to be an arbitrary constant much smaller than the typical
(r.m.s.) fluctuation of $h({\bf x})$.
The value of $\Delta$ does not affect our exponents 
and we need to consider it explicitly only in the arguments 
of Sec.~\ref{Hyperscaling}; in other places we may implicitly 
scale $h({\bf x})$ such that $\Delta=1$. 
In STM images of rough
metal surfaces $\Delta$ is usually the height of a single step on the surface.

The contour plot consists  of closed nonintersecting lines in the  
plane that connect points of equal height, which we call
{\em contour loops} (see Fig.\ref{fig_loops}, below). 
Every random-surface configuration  maps to a configuration of 
contour-loops; when the probability weights of the respective 
configurations are taken into account, this defines a mapping of the
random-surface ensemble, to the {\em contour-loop ensemble}. 
The contour loop ensemble arising from self-affine random surfaces 
is (we shall argue) self-similar; the loops are connected clusters 
that can be studied using 
scaling, just as (critical) percolation clusters have been analyzed
in previous work\cite{stauffer_book}.

For every contour loop in the loop ensemble we define a  loop length $s$ and a 
loop radius $R$. In all the 
examples we study the heights are defined on an $L\times L$  
square lattice with  lattice constant $a$. The loop length is measured with 
a ruler of length $a$ while the loop radius (really a diameter)
is defined as the side of the smallest box that completely 
covers the loop; see \Fig{fig_walks}.

In the loop ensemble we define a joint distribution $\tilde{n}(s,R)$ 
(independent of the contour spacing $\Delta$)
such that the number of loops with length in $(s,s+ds)$ and
radius in $(R,R+dR)$, per unit area, is 
\be{nsal-def}
      \Delta^{-1} \tilde{n}(s,R)ds~ dR . 
\ee
The factor $\Delta^{-1}$ has the obvious significance that
if one halves the contour spacing, one has twice as many contours. 

Assuming that the loop ensemble is scale invariant, we expect that 
$\tilde{n}(s,R)$ has a  scaling form 
\be{nsal}
\tilde{n}(s,R) \sim s^{-y} f_n(s/R^{D_f}) \ .
\ee
Here $D_f$ is the  fractal dimension, and $y$ is simply  
related to the length distribution exponent   
$\tau$, which we define in the next section.
  
In practice the exponent $D_f$ is measured by the scaling relation
\begin{equation}
\label{dimdef}
 \langle s \rangle (R) \sim R^{D_f} \; , 
\end{equation}
where 
\be{sbar_def}
 \langle s \rangle (R)\equiv  \int _R ^{R+\delta R} s  \tilde{n}(s,R)ds dR/ 
 \int _R ^{R+\delta R} \tilde{n}(s,R)ds dR 
\ee
is the average loop length for loops whose radius falls in
the interval $(R, R+\delta R)$, $\delta R \ll R$. 
The scaling in Eq.~(\ref{dimdef}) follows immediately
from the assumed scaling form in Eq.~(\ref{nsal}).

The dimension defined in \Eq{dimdef} is really the {\it scaling} dimension 
of the loop length, i.e., it defines the relation between bigger and
smaller loops in the distribution. 
On the other hand, the {\it proper} fractal dimension 
(either the Hausdorff dimension $D_H$ or the self-similarity dimension) 
refers to the relation between bigger and smaller pieces of the same loop.
Thus $D_H$ is defined by $s \sim a^{-D_H}$, 
i.e. how the loop length scales with the ruler size. 
When the contour-loop  distribution is self similar 
(as we shall assume), the two kinds of dimensions are equivalent.

\subsection {Loop length distribution exponent}
\label{Length-dist}

We define the loop number density $\tilde{P}(s)$ 
so that $\Delta^{-1} \tilde{P}(s) ds$
is the {\em total} number of loops, per unit area (measured in sites), 
with lengths in $(s,s+ds)$;
a  related 
distribution of loop lengths, $P(s)$, is defined such that
$\Delta^{-1} P(s)ds$ is the number of loops passing  
{\em through a fixed point} (say the origin) with lengths in the
range $(s,s+ds)$. In lattice models (including our numerical
examples in Sections \ref{sec_gauss}, \ref{sec_SSM}, and \ref{sec_exp}), 
$s$ is an integer and $P(s)$ is essentially the probability that the
loop has length $s$. 

From comparison to \Eq{nsal-def} it is obvious that  
\be{P-ns}
 \tilde{P}(s)=  \int_0 ^ \infty ~ \tilde{n}(s,R)  dR.
\ee
Since the total number of sites along a loop is equal  
to its length $s$ we have 
\be{tildeP}
 P(s) = s \tilde{P}(s);
\ee
the additional factor of $s$ is because each site could be 
the origin in the definition of $P(s)$.

Assuming that the loop ensemble is scale invariant  we can define
the length distribution exponent $\tau$ by
\begin{equation}
\label{taudef}
 P(s) \sim s^{-(\tau-1)}; \quad
\tilde{P}(s) \sim s^{-\tau}.
\end{equation}
This is to hold for large contour loops, i.e. those of radius much bigger then 
the microscopic scale $a$. 
Indeed, inserting \Eq{nsal} into \Eq{P-ns} gives \Eq{taudef}, with
\be{tau-y}
  y=\tau + 1/D_f \ . 
\ee

On the other hand, we could also define $\tilde{n}(R)$ such that
$\Delta ^{-1} \tilde{n}(R) dR $  is the total number of loops, per unit area, 
whose radius is in the range $(R,R+dR)$. 
Obviously 
\be{nR-ns}
\tilde{n}(R)=  \int_0 ^ \infty ~ \tilde{n}(s,R)  ds.
\ee
Doing the integral and then eliminating $y$ using (\ref{tau-y}) gives 
\be{nR-tau}
\tilde{n}(R)\sim 1/R^{1+D_f(\tau-1)}
\ee
We would have obtained the same result more quickly (and more dubiously)
had  we assumed a strict relationship between radius and length, 
$s= (const) R^{D_f}$, rather than write (\ref{nsal}). 

\subsection {Loop correlation function}
\label{Loopcorr}

The  loop  correlation function 
$G({\bf r})$ measures the probability that two points 
separated by ${\bf r}$ lie on the same contour loop. This correlation 
function is non-local, for the connectedness of the two points
depends on every site on the portion of loop between them.
This {\it loop} correlation function should be 
distinguished from the 
{\it level-set} correlation function which simply measures
the probability that two points separated by ${\bf r}$ are at the same 
height.  For the loop correlation function to be well defined 
the contour lines
are considered to be of finite width given by the microscopic scale  $a$. 
Due to  rotational symmetry of the loop ensemble,
$G({\bf r})$ depends on $r=|{\bf r}|$ only, and for large separations
($r \gg a$) we expect it to fall off as a power law:
\begin{equation}
\label{loopcorr}
  G(r) \sim \frac{1}{r^{2 x_l}} \; .
\end{equation} 
This equation defines the loop correlation exponent $x_l$ which is at the 
heart of the scaling theory of contour loops developed below.

\section{Scaling relations}
\label{sec_scaling}

In this section  we derive
scaling relations among the roughness 
exponent $\alpha$, and the three
geometrical exponents --  $D_f$, $\tau$, and $x_l$ -- 
associated with contour-loops 
and defined in Sec.~\ref{sec_measures}.
These formulas are corollaries of the
self-affineness of the rough surface, \Eq{affinedef}. 
Furthermore, for growth on an initially flat substrate
the heights will be uncorrelated beyond a certain time-dependent length scale
and the large contour loops are best modeled as hulls of percolation
clusters. This  implies a crossover to a different set of exponents
as worked out in Sec.~\ref{subsec_perc}.
The scaling relations -- including the finite-size and finite-time
forms in Sec.~\ref{subsec_FSS} and Sec.~\ref{subsec_finitet} -- 
will serve as a useful tool
for analyzing the surface morphologies obtained from numerical 
simulations and in experiments 
(see sections~\ref{sec_SSM} and \ref{sec_exp}). 

There are three stages of the main derivation. First, we establish a
relationship between the self-affine exponent $\alpha$
and the loop-size distribution exponent $\tau$; it is analogous to
the hyperscaling relation among percolation exponents. 
Second, we find a sum rule (analogous to the susceptibility sum rule)
relating the loop correlation exponent $x_l$
of Section \ref{Loopcorr} and the loop-size distribution exponent $\tau$.
Third, we present a conjecture that the 
loop correlation exponent has a value  $x_l=1/2$, 
which is super-universal in the sense that it is independent of $\alpha$. 
(This conjecture is supported by an exact calculation of $x_l$ in the 
extreme cases, i.e. $\alpha=0$ (equilibrium rough case) and $\alpha=1$.)
Finally, these relations taken together yield formulas for
$D_f$ and $\tau$ (\Eq{Df-alpha}) as a function of $\alpha$. 

\subsection {Hyperscaling relation}
\label{Hyperscaling}

If we parametrize a loop as ${\bf l}(s)$, where $s$ is the arc length
as measured by a ruler of length $a$, then 
after the rescaling given by Eq.~(\ref{affinedef}) it 
is mapped to,
\begin{equation}
\label{yscaling}
 {\bf l}(s) \rightarrow b^{-1} {\bf l}(b^{D_f}s) \; .
\end{equation}
This scaling property  of the contour ensemble  justifies 
the power law dependence of $G(r)$ on $r$ and $P(s)$ on $s$, 
in Eq.~(\ref{loopcorr}) and Eq.~(\ref{taudef}) respectively.

In writing Eq.~(\ref{yscaling}), we made a nontrivial hypothesis 
that the contours of the height function obtained by 
coarse-graining a given realization of
$h({\bf x})$ are statistically the same as the coarse-grained version 
of the contours of $h({\bf x})$.
We know of no coarse-graining procedure for the height function which assures
that the contours will stay the same. It will happen that, 
near a saddle-point of $h({\bf x})$, two loops (both of height $h_{\rm lev}$)
approach closely, but the coarse-graining shifts the height of the saddle-point
across $h_{\rm lev}$ so that the coarse-grained versions of the loops coalesce
into one loop. Whether this phenomenon makes a relevant contribution to
our scaling relations depends on the frequency of close 
approaches\cite{FN_water}. 

To determine the scaling of $\tilde{n}(R)$ first apply
the rescaling Eq.~(\ref{affinedef}) to each configuration of $h({\bf r})$;
this maps the contour ensemble to a
new contour ensemble with rescaled contour interval
$\Delta'=b^{-\alpha} \Delta$.
The {\it total} number of contours with radii in the range
$(R,R+dR)$, in a box of side $L$, is
$L^2 \Delta^{-1} \tilde{n}(R) ~ d R$, by our definition in
Sec.~\ref{Length-dist}. 
Since the contours are mapped 1-to-1 
(according to the hypothesis in Eq.~(\ref{yscaling})),
we can equate this with the number of new contours in a
box of side $L/b$ and of radius in $(R/b, R/b+dR/b)$,  
which is $(L/b)^2 \Delta'^{-1} \tilde{n}'(R/b) dR'/b$.
On the other hand, by self-affineness the new height ensemble, 
is statistically identical to the original
one (for large $R$); this holds as well for the new contour ensemble, 
thus $\tilde{n}'(R)\equiv \tilde{n}(R)$. 
So we obtain $\tilde{n}(R/b)  = b^{3-\alpha} \tilde{n}(R)$, which implies
the scaling behavior
\begin{equation}
\label{etaR}  
 \tilde{n}(R) \sim R^{-3 + \alpha} \ . 
\end{equation}

Equating Eq.~(\ref{etaR}) and Eq.~(\ref{nR-tau}) leads to the first 
scaling relation (called ``hyperscaling'')
\begin{equation}
\label{scaling2}
  D_f \; (\tau - 1)  = 2 - \alpha \; .
\end{equation}
This scaling relation has been derived previously by Huber 
{\em et al.}\cite{greg_paper} in a slightly different context, and 
in a somewhat different form by Isichenko and Kalda\cite{kalda}.   
Unlike the usual hyperscaling relation for percolation clusters which 
can be derived from the assumption that the number of large clusters
does not grow with scale of observation, here that number grows as 
a power with exponent $\alpha$\cite{aizenman}.

\subsection {Sum rule}

A second scaling relation can be derived from a sum rule.
To start off, let's separately consider the loop correlation function
for different loop sizes $s$. 
Let $G_s({\bf r})$  be the  probability that point ${\bf x}+{\bf r}$ is on the
same loop as ${\bf x}$, given that the loop has length $s$. 
In light of the self-similarity of the loop ensemble, 
it is reasonable to assume
  \be{Gsform}
      G_s({\bf r}) \sim s^{m} |{\bf r}|^{-a} f_{Gs}(r/s^{1/D_f}) \ ,
  \ee
where $m$ and $a$ are as-yet undetermined exponents, and $f_{Gs}()$
is a scaling function. 
The reason we must scale $r$ by $s^{1/D_f}$ is that this is the 
typical diameter $R$ of the loop ($s\sim R^{D_f}$). 

Now, the sum of $G_s({\bf r})$ over all lattice points is the
expectation of the total number of points in the loop, which was
given to be $s$, hence (substituting from (\ref{Gsform})) \cite{FN-chiG}
   \be{Gsdr}
       s =  \int d^2{\bf r} G_s({\bf r}) \sim s^{(2-a)/D_f+m}
   \ee
which gives one relation between the exponents $a$ and $m$ introduced
in (\ref{Gsform}):
   \be{Gsexptrel}
           2-a = D_f(1-m)
   \ee

On the other hand, the total loop correlation is the integral
of $G_s({\bf r})$ over the loop distribution function
$P(s)$ given by \Eq{taudef},  thus
   \be{Gssum}
           G({\bf r}) = \int ds P(s) G_s({\bf r})  \sim 
            r^{-a} (r^{D_f})^{m+2-\tau}  
            =  r^{D_f(3-\tau)-2}
   \ee
where (\ref{Gsexptrel}) was used to eliminate {\it both}
$a$ and $m$ in the result.  Equating the exponent of $G({\bf r})$ in
(\ref{Gssum}) to the one defined by (\ref{loopcorr}), we obtain the
scaling relation
\begin{equation}
\label{scaling1}
 D_f (3 - \tau)  = 2 - 2 x_l \; .
\end{equation}

The above scaling relations, 
Eq.~(\ref{scaling1}) and Eq.~(\ref{scaling2}), can be combined into 
expressions 
(which were originally presented in Ref.~\onlinecite{loop-PRL})
for the fractal 
dimension $D_f$ and the exponent $\tau$:
\be{results1}
    D_f  = 2 - x_l - \alpha/2 
\ee
\be{results2}
    \tau - 1 =  \frac{2 - \alpha}{2 - x_l - \alpha/2} \; .
\ee

The first scaling relation is reminiscent of the relation 
\begin{equation}
\label{man}
D = 2 - \alpha
\end{equation}
due to Mandelbrot\cite{mandelbrot}. The important difference is that 
\Eq{man} gives the fractal dimension $D$ of the {\em level set}  of 
a random self-affine surface, and {\em not} the fractal
dimension of a single contour loop. (We emphasize this point because there 
has been some confusion in the literature where the two dimensions have 
been  equated.)  

Olami and Zeitak\cite{olami} considered the same loop ensemble, but
mostly focused their attention on the ``islands'' contained in the loops 
rather than the contours; their ``$\tau$'' exponent (which we call
$\tau_{\rm ZO}$) refers to the distribution of island sizes. 
They derived a formula $\tau_{\rm ZO}= 2 -\alpha/2$ (in our notation). 
It is easy to show  $2 (\tau_{\rm ZO}-1) = D_f(\tau-1)$ --
the ``2'' here is the fractal dimension of these islands\cite{olami};
upon inserting this conversion, their formula turns out to say
$D_f(\tau-1) = 2-\alpha$, which is the same as our \Eq{results2}.

\subsection {Loop correlation exponent}
\label{sec_loopexp}

Now we turn our attention to the contour correlation exponent, and  
we {\em conjecture} that  
\be {loopcorr-conj}
     x_l = 1/2 
\ee 
is super-universal in that it is {\em independent} of $\alpha$. 

In the case of an $\alpha=0$ Gaussian surface, we know $x_l=1/2$ exactly 
for a solvable statistical-mechanics model of contour loops, equivalent
to the critical $O(2)$ loop model on the honeycomb lattice\cite{nien_rev}. 
Details are in Appendix \ref{app_O2loop}.
By invoking universality this is valid for
all logarithmically rough random Gaussian surfaces.

The exact value of $x_l$ can also be determined for $\alpha=1$. Namely,   
the fractal dimension ($D_f$) of a 
contour loop must satisfy  $D_f\leq D$  
since it is a subset of the level set, which has dimension $D=2-\alpha=1$;  
\Eq{man}. 
On the other hand $D_f \geq 1$
since a loop has topological dimension one.  From these inequalities  
we conclude that for
$\alpha = 1$ the fractal dimension of a contour loop is
 $D_f = 1$. This  in turn leads to $x_l = 1/2$, from 
Eq.~(\ref{results1}).

The validity 
of conjecture (\ref{loopcorr-conj}) for general $\alpha$ 
has been checked, to date, only
through the numerical simulations reported in Section~\ref{sec_gauss} 
and in numerical simulations of 
Zeng {\em et al.}\cite{chen_disorder}.   

Since $x_l = 1/2$ for $\alpha=0$ and $\alpha=1$, a proof of monotonicity 
of $x_l$ with $\alpha$   would suffice to establish the conjecture.  
Even that is very difficult owing to the non-local definition of the 
loop correlation function.

\subsection {Combined scaling relations}
  
Equipped with the (super-universal)
conjectured value of the loop exponent $x_l=1/2$,
and the scaling relations, Eqs. (\ref{results1}) and (\ref{results2}), 
we find the following formulas for the geometrical exponents of  
contour loops of a self-affine surface with roughness exponent $\alpha$:
\bea{Df-alpha}
    D_f  & = & \frac{3 - \alpha}{2}  \nonumber \\
    \tau - 1 & =  & \frac{4 - 2\alpha}{3 - \alpha} \; .
\eea
These relations form the basis of the contour loop analysis 
of rough surfaces, which we implement in the following sections.

Our formula for $D_f$ 
differs from the one proposed by Isichenko \cite{isich} 
\be{isichdim}
D_f^{\rm Isichenko} = \frac{10 - 3\alpha}{7} \; , 
\ee
which was derived from an approximate ``multiscale" analysis. We note that 
the formula for $D_f$ in \Eq{isichdim} gives the wrong result in the 
$\alpha=0$ case, where $D_f=3/2$ is exact\cite{saleur}.

\subsubsection{Finite-size scaling}
\label{subsec_FSS}

For realistic rough surfaces the self-affine scaling will be cut off at 
large lengths either by the correlation length or the system size. 

In the case that self-affine scaling is cut off only by the 
system size $L$,  we can extend the power laws derived above
for the average loop length,
the size distribution of loops, and the loop correlation function, 
into scaling forms:
\bea{FSSform}
  \langle s \rangle (R,L) & = & R^{D_f} f_s(R/L) \nonumber \\
  P(s,L) & = & s^{-(\tau-1)} f_P(s/L^{D_f}) \nonumber \\
  G({\bf r},L) & = & |{\bf r}|^{-2 x_l} f(|{\bf r}|/L) \ .  
\eea
In the case that the self affine scaling is cut-off by a finite 
correlation length $\xi(t)<L$, our three contour-loop measures 
will display crossover effects to a different set of power laws;
we turn to this problem next.

\subsection {Percolation crossover}
\label{subsec_perc}

For a surface roughened by growth 
self affine scaling is expected to hold 
only up to a finite correlation length 
$\xi(t)$ growing with time as \Eq{dynamical}. 
At early enough stages of growth (i.e. while $\xi(t)<L$)  
the statistics at scales beyond $\xi(t)$ depend on the initial state.
Then the contour loops of the 
surface will also exhibit crossover behavior where loops whose linear size (as
measured by the radius $R$) is less than the correlation length will scale
according to the formulas derived above, while the large loops will 
exhibit scaling with percolation exponents.   

Say the initial surface is flat  (which we assume henceforth). 
Then it turns out (see Appendix \ref{app_perc}) that 
the contour loops at scale $R>\xi(t)$ are boundaries of percolation clusters. 
Although this new contour loop ensemble corresponds to a {\it non}-self-affine 
surface, it still exhibits scaling
and we derive its three loop exponents in terms of known 
percolation exponents. 
In some cases, it turns out,  the exponent values from the percolation regime
and from the self-affine surface are not so different; thus a careless
analysis might yield spurious exponents. 

It is easy to see that at $r>\xi$, we can model the actual heights (not 
height differences)  as statistically independent, since
(by definition of the correlation length)
the distance $\xi(t)$ is the farthest that an event 
can influence another in time $t$. 

In appendix \ref{app_perc}, we derive the geometrical exponents
   \be{perc_results}
   D_{f,p} =7/4=1.75 , \ \tau_{p}=18/7=2.571, \ 2 x_{l,p}=5/4 =1.25
   \ee
which apply to loops at scales larger than $\xi(t)$
(the percolation-regime scaling). 
These exponents are the same for any $\alpha$. 
An important corollary is that
power-law scaling in the loop analysis is 
{\it not} necessarily a signature of self-affine behavior. 
Indeed, most real surfaces never reach a clear self-affine regime, hence
their loops are probably in the percolation regime.

\subsubsection { Finite-time crossover scaling forms}
\label{subsec_finitet}

The complete crossover between the self-affine and percolation regimes
is described by scaling forms parallel to (\ref{FSSform}). 
First consider the height structure factor. 
Since,  as noted above, the heights are independent, 
their (spatial) power spectrum is flat in Fourier space:
$S({\bf q}) \sim {\rm const}$, for  $|\q |< 1/\xi(t)$; 
on the other hand, for
$|\q |> 1/\xi(t)$ the surface has already developed 
a self-affine state so \Eq{hqvar} does hold. 
The two behaviors should be combined via a scaling function $f_S()$:
  \begin{equation}
      S({\bf q};t) = |\q |^{-2(1+\alpha)}
      f_S(\q t^{1/z});
      \label{hcross}
  \end{equation}
see Fig.~\ref{fig_perccross}(a).

Thus,  at times $t$ such that $\xi(t)\ll L$, 
the behaviors (\ref{dimdef}), (\ref{taudef}), and (\ref{loopcorr})
are generalized to
  \bea{scross} 
      \langle s \rangle (R;t) & =  & R^{D_f} f_{sp}(R /\xi(t)) \nonumber \\
     P(s;t)   & =  &  s^{-(\tau-1)} f_{Pp} (s/\xi(t)^{D_f}) \nonumber \\ 
     G(r)     & =  &  r^{-2x_l} f_{Gp}(r/\xi(t)) \ .
  \eea
In each case, the scaling function is unity for argument zero, 
while for argument large it scales as a power law needed to 
give the correct exponent for the percolation regime, as 
calculated in Appendix \ref{perc-union}. 
The $t$ dependence for the prefactor of each 
percolation-regime power law
is given by the requirement to patch the above two dependences together 
when the scaling-function argument is of order unity. 

Figures \ref{fig_perccross} (b)--(d)
illustrate the shapes of the three loop measures.
Notice that the ``knee'' around $r=\xi(t)$ appears
more strikingly in the Fourier analysis
than in any of the loop analyses.
Although the percolation-regime and self-affine exponents have
fairly similar values, 
the difference grows larger as $\alpha$ gets larger.
The crossover is evident in our simulated Gaussian data (see 
\Fig{fig:gauss04_cut2}).


\section{Simulation: Gaussian random surfaces}
\label{sec_gauss}

Here we test the validity of our scaling relations and the 
effectiveness of determining $\alpha$ from contour loops,  
under the controlled circumstances 
provided by computer generated surfaces with known $\alpha$. 
The  surfaces we construct are self-affine with 
Gaussian fluctuations of the height.

\subsection {Construction}
\label{sec_gauss_construct}

Random Gaussian surfaces are generated numerically as an $L/a\times L/a$
matrix $h({\bf x})$ 
of real-valued  heights associated with the vertices $\{{\bf x}\}$ 
of a square lattice of size $L$, with lattice constant $a$. 
A particular realization of $h({\bf x})$ is given by 
Fourier transforming $\tilde{h}({\bf q})$ 
where the wave-vectors ${\bf q}$ take their values in the 
first Brillouin zone $[-\pi/a, \pi/a] \times [\pi/a, \pi/a]$. Each 
Fourier component  $\tilde {h}({\bf q})$ is an independent
Gaussian random variable with a ${\bf q}$-dependent variance given by 
\be{hqvar-2}
 \langle |\tilde{h}({\bf q})|^2 \rangle = \frac{1}{({\bf q}^2)^{1+\alpha}} \ .
\ee
For $0\geq\alpha\geq 1$ surfaces generated in this way  
are self-affine and rough,  with a  roughness exponent $\alpha$. 

The $\alpha=0$ case of random Gaussian surfaces 
is familiar as: (i) the equilibrium-rough surface
(compare Eq.~(\ref{eq-rough})), (ii)
the surface in the Edwards-Wilkinson model, 
and (iii) the Coulomb gas representation of 
a two-dimensional critical model \cite{nien_rev} 
(see appendix \ref{app_O2loop}).
The case $\alpha = 1$ appears in
the Mullins-Herring  (diffusive relaxation) model
($\alpha=1$) of  non-equilibrium surface growth \cite{plish}.

\subsubsection {Comparison to other studies}

A popular algorithm for generating self-affine surfaces
is ``random midpoint displacement, with random successive addition''
\cite{meakin_invasion,schmitt}. 
This method iterates a step in which, starting with a self-affine surface on
a coarse grid of lattice constant $2a$, one generates heights
on a new grid of lattice constant $a$ by interpolation, and then 
adds to them random increments proportional to $a^\alpha$. 
Such an ensemble need not be Gaussian or have up-down symmetry, 
but commonly does\cite{schmitt}. We note that (i) the variance
of a site's height (relative to the initial flat surface) depends on
what iteration that site appeared, i.e. 
on how many times 2 can be divided into the site coordinates;
(ii) height-difference correlations do not always grow with distance
(they are smaller between two sites that appeared in early iterations)
and they have the anisotropy of the lattice even at large distances.
We believe our Fourier construction of self-affine surfaces
(Sec.~\ref{sec_gauss_construct})
is preferable because the resulting ensemble is
(i) spatially homogeneous and 
(ii) isotropic, on scales beyond a couple of lattice constants. 

\subsection{Curvature measurements}

We measured moments $\langle C_b^m \rangle$ ($m=2,3,4$) of the scale-dependent 
curvature, as defined in Sec.~\ref{subsec_curv}, 
for Gaussian surfaces generated by the Fourier method
described above. This data (in \Fig{fig:curv_gauss}) is a kind of 
check on the Fourier method since the mean over an infinite 
number of samples can be computed analytically. 

Self-affine scaling is evident 
on the log-log plot of the even moments in \Fig{fig:curv_gauss} (upper plot).
The roughness exponent $\alpha$ is 
obtained as the slope of a straight-line fit to the
$C_b^2$ plot, as shown in Table~\ref{tab_gauss}.
Ideally, the slopes of the $C_b^2$ and $C_b^4$ log-log plots
should be $2\alpha$ and $4\alpha$ with exactly the input 
$\alpha$ values used in constructing the random surfaces. 
This is spoiled somewhat in practice by discrete-lattice effects for
$b \leq 3 $ and by finite-size effects when $b > L/4$. 
Furthermore, $\langle C_b^4 \rangle/ \langle C_b^2 \rangle ^2$ 
should be exactly 3 for every $b$ value, even those 
for which the power-law dependence on $b$ fails, 
since this is true for any Gaussian random variable.
Indeed, the measured ratio is close to 3. 

The third moment of $C_b$ is shown in  \Fig{fig:curv_gauss}.  
Independent of $\alpha$, $\langle C_b^3\rangle$ is roughly 
zero, as expected for a random  Gaussian surface
which posses a  $h\to -h$ symmetry 
(i.e., the valley bottoms and the hill tops are
equivalent for a Gaussian surface).

\subsection{Loop measurements} 
\label{sec_loopmeas}

The primary motivation for our Gaussian surface simulations was 
an initial test of the scaling predictions for the contour loop
exponents from Sec.~\ref{sec_scaling}. 
A contour plot of a sample surface configuration for 
$\alpha=0.4$ is shown in \Fig{fig_loops}. (A similar plot for
$\alpha=0$ was published in Ref.~\onlinecite{loop-PRL}.)

\subsubsection {Measurement procedure}

In a single run, which would typically take 10 minutes on an Sun Sparc5 
workstation,  
25 surfaces of specified roughness $\alpha$ were generated.  
For each  surface typically 400 points were chosen at random, 
and through each point a contour loop was constructed using the loop 
finding algorithm as explained in Appendix~\ref{app_loopalg}. While 
each loop was being traced points along the loop were used 
to evaluate the loop correlation function $G(r)$. 
For each contour loop its radius and 
length were measured and used to determine the length  distribution of 
contour loops ($P(s)$) and the average loop length $\langle s \rangle$
as a function of the loop radius $R$. 

\subsubsection{Results}

In order to measure the geometrical exponents $D_f$, $\tau$, and $x_l$ we
plotted the data for system size $L=512$ on a log-log graph and 
performed least-squares linear fits. 
Data was selected for fitting 
from the range in which a well developed power law was observed; see 
Figures \ref{Fig:Sr_512}, 
\ref{Fig:PS_512},  and 
\ref{Fig:Gr_512}.
The results are given in Table~\ref{tab_gauss}. We
find excellent agreement between the predictions of the scaling theory and
the measured geometrical exponents. In particular, note that the simulations
confirm the super-universal nature of the loop correlation exponent 
$x_l=1/2$.   

The loop correlation function $G(r)$ has a 
a size dependence which biases a direct  fit to the exponent $2x_l$;
finite-size scaling (see below) partially overcomes this systematic error.  
Our theory (Sec.~\ref{sec_loopexp}) indicates that 
$G(r)$ has a universal {\it exponent} 
$2x_l=1$; in fact, as  shown in \Fig{Fig:Gr_512}, 
$G(r)$ itself appears practically independent of $\alpha$. 
Closer examination reveals that the coefficient in $G(r) \sim 1/r$
decreases slightly as $\alpha$ grows.
Furthermore, the fitted values of $2x_l$ (see Table \ref{tab_gauss})
decrease a bit with $\alpha$, which we attribute to 
the systematic error just mentioned, combined with 
the small $\alpha$-dependence of the shape of 
the ``knee'' in the finite-size behavior of \Fig{Fig:FSS_Gr_04}.
There is no indication in the extracted $2x_l$ values of 
any non-monotonic dependence on $\alpha$;  
as shown in section \ref{sec_loopexp}, monotonicity of $x_l(\alpha)$ 
is sufficient to prove $2x_l=1$, independent of $\alpha$.  

A better measure of the geometrical exponents was obtained from 
a finite size scaling analysis of the data. 
Using the scaling forms in \Eq{FSSform} we produced data collapses
(``scaling plots''). 
Sample data for the $\alpha=0.4$ case are given in
Figs.~\ref{Fig:FSS_PS_04}(a) and \ref{Fig:FSS_Gr_04}(a);
the data collapse is shown in Figs.~\ref{Fig:FSS_PS_04}(b) and
\ref{Fig:FSS_Gr_04}(b). 
From the loop-size distribution plots like \Fig{Fig:FSS_PS_04}(b), we 
extracted both the exponents $D_f$ and $\tau-2$.
Similarly, we obtained $2x_l$ from the loop correlation function 
\Fig{Fig:FSS_Gr_04}(b); in this case we don't fit another exponent
since $r$ obviously scales as $L^1$.
(We did not carry out finite-size scaling of the 
$\langle s \rangle$ versus $R$ plots such as \Fig{Fig:Sr_512}, since
there was no obvious change in the slope as a function of $R/L$.)
The geometrical exponents giving the best data collapses
are reported in the ``FSS'' columns of Table~\ref{tab_gauss}. 
The reported uncertainties were estimated by the interval 
over which changes in the exponent value  did not visibly 
worsen the data collapse. 

Note in Table~\ref{tab_gauss} how the finite-size scaling exponents 
agree better with the scaling theory of Sec.~\ref{sec_scaling}  
than the exponents obtained from ``direct'' fitting of the  data 
to power laws. 
The discrepancy becomes more obvious at larger values of 
$\alpha$.   
We infer from this that Gaussian surfaces with a large value of the 
roughness have more pronounced finite size effects which lead to an 
overestimate of $D_f$ and $\tau$. This is  of relevance to 
experimental data where the system size is typically 
not a tunable parameter, and the geometrical exponents are necessarily  
measured using the direct-fit method.

\subsubsection {Relation to a previous simulation}

Numerical measurements of the fractal dimension of contour loops have been 
done  by Avellaneda {\em et al.} \cite{avella}.
They  found $D_f = 1.28 \pm 0.015$ for an $\alpha= 0.5$
surface, which  is close to the predicted value $D_f = 1.25$,
from Eq.~(\ref{Df-alpha}). They also measured the combination 
$D_f(\tau-2)$ (their ``$\alpha$'') which describes the 
scaling of the probability that a loop passing through a fixed 
point has a radius larger than $\rho$, with $\rho$. (We evaluate this 
quantity by integrating $n(s,R)=s\tilde{n}(s,R)$, from \Eq{nsal}, 
over all $s$ and for $R>\rho$). The numerical 
result they quote, $D_f(\tau-2)=0.21 \pm 0.017$, is in fair agreement 
with our prediction $D_f(\tau-2)=(1-\alpha)/2=0.25$ for $\alpha=0.5$, 
which follows from \Eq{Df-alpha}.

Wagner {\it et al}\cite{meakin_invasion}
simulated a form of invasion percolation where the
threshold pressures have the form of a self-affine surface.
Hence the perimeters of the invaded clusters are the same
as the contour lines of the surface. They claim that the
perimeter dimension is ``consistent'' with Isichenko's formula, 
our \Eq{isichdim}, but do not quote an error; perhaps their
precision was such that the prediction of (\ref{isichdim})
could not have been distinguished numerically from the one 
we believe to be correct, (\ref{Df-alpha}). 

Ref.~\onlinecite{meakin_invasion}
also mention measuring a behavior $r^{-\gamma}$ with 
$\gamma\approx 0.9$
for  the correlation between successive filled sites.
If this were simply a correlation of 
two randomly chosen points along a perimeter, it would be
identical to our 
loop correlation functions, 
$G(r)$ or $G_s(r)$ defined in (\ref{loopcorr}) or 
(\ref{Gsform}); in fact, the filling process would appear to
depend on correlations of the surface gradient and might
have a somewhat different exponent.

\subsection{Surfaces with a finite correlation length}
\label{subsec_gauss_finitecorr}

To test the  percolation analysis of self-affine rough surfaces
with a cutoff, as derived in Appendix \ref{app_perc}
and summarized in Sec.~\ref{subsec_perc}, 
we performed curvature and loop measurements on Gaussian surfaces 
with a correlation length $\xi$.  
The correlation length is incorporated in the Fourier method of 
generating Gaussian surfaces by changing the variance of 
$\tilde{h}({\bf q})$ in \Eq{hqvar-2} to: 
\be{hq_var_cut} 
\langle |\tilde{h}({\bf q})|^2 \rangle = \left\{ \begin{array}{ll}
      |{\bf q}|^{-2(1+\alpha)} & \mbox{for $|{\bf q}| > \pi/\xi_q$} \\
   (\pi/\xi_q)^{-2(1+\alpha)}    & \mbox{for $|{\bf q}| \leq \pi/\xi_q$}
      \end{array} \right.
\ee

The effects of the cutoff are summarized in \Fig{fig:gauss04_cut2}, 
which should be compared to the theoretical prediction  
of \Fig{fig_perccross}. The curvature and loop data shown in the figure 
are for system size $L=512$.  

The second  moment of the curvature displays self-affine scaling 
with roughness $\alpha=0.4$ up to a length scale set by $\xi_q$, and
beyond this scale it levels off; see \Fig{fig:gauss04_cut2}a). 
We checked that  the third moment of the curvature  vanishes, 
as expected  since the height fluctuations are still Gaussian, while 
the fourth moment follows affine scaling up to roughly the same correlation 
length as the second moment.  

The loop measures exhibit distinct crossover behavior, as seen in 
figures \Fig{fig:gauss04_cut2} b) through d). 
For loops whose radius is smaller than the correlation length, 
which is here $\xi\approx 20$, we find values of the geometrical exponents 
consistent with those extracted previously for $\alpha=0.4$ random 
Gaussian surfaces.  For loops whose linear size exceeds the 
cutoff,  scaling consistent with the percolation analysis is found. The 
actual numerical values extracted by fitting the $20<R<200$ data to a
power law are somewhat larger than expected ($2x_l=1.46(6)$, $D_f=1.7(1)$, 
and $\tau=2.63(1)$)  which we attribute to 
finite size and/or crossover effects. To check this we also 
simulated a Gaussian surface 
with completely uncorrelated heights, i.e., with $\xi=1$ and 
system size $L=512$, for which we find (by the direct-fit method):
\be{uncor}
2x_l=1.26(3) \: ,  \  D_f=1.70(2) \: ,  \ \tau=2.565(10)
\ee
in good agreement with \Eq{perc_results}.

\section{Simulation: Non-equilibrium growth model}
\label{sec_SSM}

In this section, the linear and nonlinear roughness measures of
Sec.~\ref{sec_measures},  which in Sec.~\ref{sec_gauss}  
were tested on artificial Gaussian random surfaces,
are now applied to growth-roughened surfaces produced by 
a simple random deposition model, the well-known ``single-step model''. 
Our results are in support of the view that the single-step model
produces self-affine morphologies. 

\subsection{The single-step model}

We implemented the ``single-step model''\cite{meakin,plischke-SSM,liu}
in $d=2+1$ dimensions\cite{FN-kim}.
(More details on this model are found in Sec.~III~F of 
Ref.~\onlinecite{meakin}, or Sec.~II~A of 
Ref.~\onlinecite{liu}.) There is one control parameter $p_+$.
The allowed configurations are just those of the BCSOS model: each site of 
a square lattice has an integer-valued height and 
neighboring heights must differ by $\pm 1$.
The Monte Carlo rule is that in each time step a deposition event occurs
with probability $p_+$ or an evaporation event (inverse of a deposition
event) occurs with probability
$1-p_+$; once it is decided which type of event occurs, a site is 
picked at random among those sites at which that event is 
allowed~\cite{FN-nonlocalSSM}.

We begin by an overview of the theoretical expectations.
Up-down symmetry switches $p_+ \leftrightarrow 1-p_+$;
thus we need only report data for $0<p_+ \leq 0.5$. 
The case $p_+=0.5$ is special as  the dynamics satisfies detailed
balance. This should produce an equilibrium-rough interface,
namely the BCSOS model with all configurations weighted equally~\cite{vanB}. 
This interface, at long wavelengths,
is described by the Gaussian model of Sec.~\ref{sec_gauss}
with $\alpha=0$ (Edwards-Wilkinson behavior). 

On the other hand, the growth model for $p_+ \neq 1/2$ is believed to 
asymptotically belong to the Kardar-Parisi-Zhang (KPZ) universality class 
\cite{meakin,liu}.  
It has been proposed that $\alpha=0.4$ exactly for the 2+1 dimensional
KPZ model\cite{lassig}; however, finite-size effects,
small simulations and naive fits systematically underestimate it as 
$\alpha \approx 0.38$ \cite{family,KPZexpts}.
The KPZ behavior should be clearcut
when $p_+$ is close to $1$, but otherwise 
a crossover from
initially Gaussian to asymptotic KPZ behavior is expected, 
which will be slow
(as a function of time or system size) 
if $p_+$ is close to $1/2$. 

It turns out, in our numerical results (below), 
that $p_+=0.5$ indeed shows Gaussian behavior
and $p_+=0.1$ shows KPZ-like behavior, but $p_+=0.3$ 
consistently resembles 
$p_+=0.5$, at the sizes we could simulate
(i.e. up to $L=128$). 
We attribute this to above-mentioned crossover from initial Gaussian behavior.

\subsection {Simulations}

Starting from a flat surface, 
we ran the simulation (for systems of $128\times 128$ sites) 
for 2000 Monte Carlo steps (MCS) per site to equilibrate and then took 
data for a period of 1200 MCS/site; 
one such run took 10-15 hours of 
cpu time on a RISC-6000 workstation.
The standard-length runs (for size $L=128$) 
appeared to be insufficiently equilibrated
for $p_+=0.1$, 
since they failed to collapse on finite-size-scaling 
plots with smaller systems.
Therefore we performed one run for $L=128$, $p_+=0.1$ with 12000 MCS/site
equilibration and 10000 MCS/site for data collection;
this is the run reported in our results.
In all other cases, we believe the run time was adequate, since
much shorter runs showed no gross differences. 
We performed about four runs for each value $p_+=0.1, 0.3,$ and
$0.5$, verifying the
symmetry $p_+ \leftrightarrow 1-p_+$. 
(All measures are the
same, apart from a change in sign of $\langle C_b^3 \rangle$.) 
Only one of the $\sim 4$ runs
was selected to be fitted and plotted here; the data
sets presented as e.g. $p_+=0.1$ are actually $p_+=0.9$ in some cases.

Once every 100 MCS/site (during the data-collecting portion
of a run), we performed a measurement step on the surface.
The Fourier transform was taken of $h({\bf r})$ 
using a fast-Fourier-transform routine, but $\langle |h({\bf q})|^2\rangle$ was 
accumulated only for $\bf q$ values along 
the (1,0), (0,1), (1,1), and (1,-1) directions.
Also, in each measurement step 100 contour loops were traced out 
from random initial points,  
as described in Sec.~\ref{sec_loopmeas} and Appendix~\ref{app_loopalg}.
Statistics  were accumulated 
of the loop's radius $R$ and its number of sites (length) $s$, but not
the loop correlation function. 

\subsection {Fourier and curvature results}

The single-step model is the only one simulated in this paper
for which we evaluated Fourier spectra, which are plotted in 
\Fig{Fig:SSM_hq}; the log-log plot should have a slope
$-2(1+\alpha)$ so $\alpha$ can be extracted from a linear fit
(as in Table~\ref{tab_SSM}). Notice how the spectra are completely
isotropic with respect to the lattice directions.

The scale-dependent curvature moments were not evaluated during the runs, 
but were computed only from the final surface from each run 
(hence their statistics are much worse than for other 
measures reported here).
We computed $\langle C_b({\bf x})^m\rangle$
as defined in Sec.~\ref{subsec_curv}, for $m=2,3,4$, as a function
of the offset $b$ in the definition of $C_b$ as  a discrete Laplacian.
The results for $m=2$ and $3$ are plotted in \Fig{Fig:SSM_C}.
(All figures of the SSM are from the largest system size, $L=128$.)
\Fig{Fig:SSM_C}(a) does not show well-defined power laws. 
The curve for $p_+=0.5$ shows a smallish apparent slope, $2\alpha=0.3(1)$, 
and a downwards curvature which is plausibly consistent with the
expected logarithmic behavior, just like that of the usual 
height-difference function $D_2(b)$ (recall Eq.~(\ref{curv2}). 
The $\langle C_b^2\rangle$ curve for $p_+=0.1$ shows a larger slope, 
$2\alpha \approx 0.6(1)$, consistent with KPZ scaling. 
The $\alpha$ values in Table~\ref{tab_SSM}, 
were extracted from fits to $\langle C_b^2\rangle$ plots. 
Slopes from plots of the $\langle C_b^4\rangle$ moments (not shown) 
are consistent with $4\alpha$ for the $\alpha$ values in the table.

\subsection {Loop analysis} 

We analyzed the loop ensemble to plot the mean loop size as a function
of its radius (\Fig{Fig:SSM_SR}) and the cumulative loop-size
distribution $P_>(s)$ (\Fig{Fig:SSM_PS}). 
(Note that $s$ must have {\it even} values; thus it is necessary
to divide the nonzero values by 2 in order to properly estimate
$P(s)$, which was assumed (in Sec.~\ref{sec_measures}  and
Sec.~\ref{sec_scaling}) to be a smooth monotonic function.)

As with the Gaussian simulation of Sec.~\ref{sec_gauss}, 
the slopes of straight-line fits to log-log  plots of these data,  
give estimates of $D_f$ and $\tau-2$; they are tabulated as
``direct'' in Table~\ref{tab_SSM}. 
Alternatively, we used data like \Fig{Fig:SSM_PS}
from smaller sizes $L=32$ and $L=64$
(as well as $L=128$) to produce scaling plots analogous to 
\Fig{Fig:FSS_PS_04} (these plots not shown), extracting the
``FSS'' data in Table~\ref{tab_SSM}. 
(We do not report on $2x_l$
since we did not evaluate the loop correlation $G(r)$ 
in our SSM simulations.)

It is interesting to compare the four different measures
of $\alpha$ included in Table~\ref{tab_SSM}.  Those from
$\langle |h({\bf q}|^2 \rangle$ seem to have the smallest
statistical errors (and the most sensible values). 
The closely related $\langle C_b^2\rangle$ result is 
expected to be worse,  not only because the statistics are
poor in our implementation of the SSM simulation (see above), but also
because it uses $h({\bf x})$ values from more widely spaced $\bf x$
and is therefore more sensitive to the system size. 

The next best method seems to be the $D_f$ loop analysis from
$(\langle s\rangle ,R)$ plots;
curiously, it appears that the $D_f$ fits
show smaller run-to-run fluctuations than the
$\langle |h({\bf q}|^2 \rangle$ fits. 
As also observed in the Gaussian runs (Sec.~\ref{sec_gauss}), 
the $P_>(s)$ analysis showed more obvious finite-size effects;
direct fits to $\tau$ are unreliable and only
finite-size scaling plots give reasonable results.

\section{Analysis of experimental data}
\label{sec_exp}

In this section we test our nonlinear measures 
against experimental scanning tunneling microscopy (STM) data sets.
Rough metal surfaces grown under several 
conditions are believed to develop a morphology with self-affine 
scaling,  but only up to a time-dependent correlation length $\xi(t)$
as discussed in Sec.~\ref{subsec_perc}. 

The most detailed analysis was done for  
the vapor deposited Ag surface on a quartz substrate
of Palasantzas and Krim \cite{krim}.
We obtained a  $400\times 400$ height array corresponding to an  
STM image of a $702$nm thick Ag surface and
performed curvature and loop measurements. Note that all the results quoted
below are from a {\em single} height profile. 
All in-plane lengths will be measured in units 
of the grid of this data, which is 1.625 nm.

We also report briefly (Subsec.~\ref{subsec_otherdata})
a less thorough analysis of an STM  data set
from a different, but still self-affine, growth regime
showing KPZ scaling.

\subsection {Quadratic measures and curvature moments}

Palasantzas and Krim\cite{krim} 
originally evaluated the roughness exponent 
\be{PKresult}
\alpha = 0.82(5) \ 
\ee
from a fit to the standard (quadratic) height
correlation function $D_2({\bf r})$ defined in \Eq{struc}. 

That correlation is similar to our second curvature moment 
$\langle C_b^2 \rangle$.  This quantity shows a power 
law dependence on the scale $b$, up to a 
correlation length which was estimated to be $\xi=25(5)$;  see 
\Fig{Fig:Cb_Krim}(a). 
A linear least squares fit of the data with $b<\xi$ 
gave $2\alpha=1.7(1)$, agreeing (as expected) with (\ref{PKresult}).

The third moment $\langle C_b({\bf x})^3\rangle$, \Fig{Fig:Cb_Krim}(b),  
shows distinct non-Gaussian behavior, as expected for non-equilibrium growth.
It reaches a maximum at length scale $b\approx 23$ 
which correlates well with $\xi$. This  indicates a 
morphology consisting of grains of typical size $\xi$ that are rounded 
at the top. Such a morphology is clearly seen in three-dimensional renderings
of the STM data in Ref.~\onlinecite{krim}, or the gray-scale image
in Ref.~\onlinecite{kleban}. 

\subsubsection {Kleban's nonlinear measure}
\label{subsec_kleban}

Recently, Kleban {\em et al} 
defined a non-quadratic roughness measure rather different from
any of those mentioned in Sec.~\ref{sec_measures}. 
First, for every scale $b$ they constructed a smoothed version 
$H_b({\bf x})$ of the height function, 
as the average of $h({\bf x'})$ for ${\bf x}'$ in 
a $b \times b$ square centered at $\bf x$ (alternatively by
convolving with a Gaussian weight function of width $b$.)
Then they calculated the histogram $P(H_b)$ of $H_b({\bf x})$ values 
for ${\bf x}$ ranging over the entire sample, and the 
skew moment of this distribution.
(Of course, $P(H_b)$ is defined  for a rough surface
only when a finite-size or finite-time  cutoff is present.)

When applied to the STM data of Palasantzas and Krim~\cite{krim}, 
the distribution of $H_b({\bf x})$ 
{\it appears} Gaussian when $b=0$, i.e. for the raw data.
That is rather mysterious,
since the surface certainly lacks up/down symmetry: it consists of
deep, narrow crevasses and rounded hills. 
Each crevasse
should contribute to a long tail on only the $h < \overline{h}$ side of the 
distribution, while each hill contributes a peak and a sudden drop to
zero on the $h > \overline{h}$ side. 
However, the fluctuation in height from hilltop
to hilltop smears out that sharp feature, 
making a spuriously symmetric distribution.
As pointed out in Ref.~\onlinecite{kleban}, 
their smoothing of $h({\bf x})$ eliminates the deep crevasses 
so the {\it smoothed} surface $H_b({\bf x})$ {\it does} have a
skewed height distribution  (with skewness dependent on the
observation scale $b$). This is consistent with our own
conclusion that the height fluctuations are non-Gaussian.

\subsection {Contour-loop analysis}

We perform loop measurements and 
check whether the different scaling relations derived 
in section~\ref{sec_scaling} are satisfied.  
The moments of the scale dependent 
curvature are  used as an independent measurement of the roughness
and to assess the Gaussianness of the height fluctuations.  

Using the loop algorithm (appendix~\ref{app_loopalg}), 
we measure the loop radii and corresponding 
loop lengths for 1000 contour loops constructed through randomly chosen points 
on the surface; however, we did not
compute the loop correlation function.
These loop measures support the scenario that the 
surface is self-affine up to a correlation length $\xi\approx 25$. 

The average loop length is plotted 
against the loop radius in \Fig{Fig:sR_Krim}. We see a decade of power
law scaling of the length with the radius, and from a linear least squares fit
of the data to a line, for $5<R<50$,   we find
\be{PKfdim}
D_f = 1.06(2)
\ee
for the fractal dimension of contour loops. Using the formula for $D_f$, 
\Eq{Df-alpha}, we calculate $\alpha=0.88(4)$ in 
good agreement with the reported value, \Eq{PKresult}. In \Fig{Fig:sR_Krim}
the dashed line corresponds to the percolation value $D_h=1.75$; we
see that loops at scales much larger then $\xi$ show scaling consistent with 
this value. 

Finally, the number of loops whose length exceeds $s$, $P_>(s)$,  is 
plotted in \Fig{Fig:PS_Krim}. The data roughly shows two scaling 
regimes with different exponents, before it is cut-off by the system 
size. The knee occurs at loop lengths $s\approx 70$ which, from 
\Fig{Fig:sR_Krim},  corresponds to a loop radius of $20$ or so; this  
again is comparable to the length scale $\xi\approx 25$ found from the 
curvature data.
 From loops whose length is in the interval $(10,30)$ 
we extract the  exponent
\be{PKTau}
\tau-2 = 0.069(5)
\ee 
while larger loops exhibit scaling consistent with the percolation value
(indicated by the dashed line). Using \Eq{Df-alpha} we find $\alpha=0.85(1)$
again in good agreement with the self-affine exponent reported by 
Palasantzas and Krim. 

To summarize, the two loop measures we evaluated, as well as 
quadratic scale-dependent curvature, all
indicate self-affine scaling with a roughness
exponent $\alpha\approx 0.85$, up to a length scale $\xi\approx 25$.
Beyond this scale the height fluctuations appear to be uncorrelated.

\subsection {Other data sets}
\label{subsec_otherdata}

The large value of $\alpha$ found for the silver-on-quartz STM data 
of Palasantzas and Krim is indicative of MBE-type growth which has 
surface diffusion as a dominant relaxation process. This motivates
the study of other data sets which might correspond to different 
universality classes of growth. For example, the KPZ  
equation describes growth dominated by desorption 
and/or vacancy formation, both of which are relaxation processes
that do not conserve particle number \cite{krimrev,krugrev}.  

Loop measurements were carried out previously  on gold 
electro--deposits by Gomez-Rodriguez {\em et al} \cite{expdep}. 
These authors suggested the fractal dimension of contour loops as 
a useful measure for characterizing the surface morphology. What was 
lacking in their analysis was an equation  relating
$D_f$ to the roughness exponent $\alpha$.
 From STM images of deposits grown in the fast and slow regime they 
determined the fractal dimension to be $D_f \approx 1.5$ and 
$D_f \approx 1.3$, respectively. 
Now using \Eq{Df-alpha} we  calculate the roughness in these two regimes
to be: $\alpha \approx 0$ and $\alpha \approx 0.4$. The first
is expected for Edwards--Wilkinson type of growth ($\alpha=0$),  while the 
second is in good agreement with the  Kardar--Parisi--Zhang 
value $\alpha=0.38$ (from most fits) 
or 0.40 (possibly exact)\cite{family,KPZexpts}.

Csahok {\em et al.}\cite{csahok} studied the surface morphology of Ni films 
vapor-deposited on a quartz substrate.
(They were interested mainly in the effects of subsequent ion sputtering 
on the film.)  We obtained an STM image of the 
as-grown Ni surface (before any sputtering)
in the form of a $256\times 256$ height array, and computed some of the
contour-loop measures for it from a collection of 10000 loops. 
The results are consistent with a (KPZ-like) 
self-affine morphology with a roughness of $\alpha=0.4$. 
Namely, the loop data 
shows a limited range of scaling for
loop radii $10<R<30$. Direct fits to a straight line of the log-log plots of
$\langle s \rangle (R)$ and $P_>(s)$ in the scaling regime,  
yield $D_f=1.31(4)$ and $\tau-2=0.22(2)$. 
After inverting the formulas for $D_f(\alpha)$ and $\tau(\alpha)$
in \Eq{Df-alpha} we obtain the estimates $\alpha=0.38(8)$ and 
$\alpha=0.44(4)$ respectively.

\section{Discussion}
\label{sec_disc}

Here we summarize our main results, compare and critique previously 
introduced measures of surface roughness, and describe some 
open problems and interesting new directions.

\subsection{Summary of results}

We introduced (in Sec.~\ref{sec_measures}) new measures for 
characterizing the spatial correlations of  rough interfaces. Their 
common property is that they are not linearly related to the structure 
function of the height. First, we introduced the scale dependent curvature. 
Its third moment is an indicator of the skewness of the 
height distribution, and thus is a good criterion for
whether or not a surface's height fluctuations are Gaussian.
Our chief focus, though, was on the ensemble of {\it contour loops} as
a novel means of characterizing surfaces. 
For a rough self-affine surface, 
the loop ensemble is critical and we introduced three kinds of 
geometrical exponent associated with it:
$x_l$ for the loop correlation function 
(probability that two points are
on the same contour loop), the fractal dimension of a contour loop, $D_f$, 
and $\tau$ associated with the length distribution of loops. 
In particular, we conjectured a {\it super-universal} value $2x_l=1$
(see Sec.~\ref{sec_loopexp}) which has been confirmed so far numerically
(e.g. in Table~\ref{tab_gauss}), but not analytically.  
The loop exponents satisfy scaling relations 
(derived in Sec.~\ref{sec_scaling}), and granting the
conjecture, their values are completely 
determined by the affine (roughness) exponent $\alpha$. 

Next, we showed how numerical values of the geometrical exponents 
can be extracted in practice from height data obtained from 
simulations or experiments. We first did this in Sec.~\ref{sec_gauss}
for artificial Gaussian surfaces (known analytically to be self-affine) 
and in Sec.~\ref{sec_SSM} for configurations from simulations
of the single-step (growth) model (believed to be self-affine); 
this served as a check to confirm the validity of our
scaling relations.
Then in Sec.~\ref{sec_exp} we processed an experimental data set -- 
an STM image of a growth roughened silver film\cite{krim} --
in the same fashion. The results here  also confirmed 
the scaling relations which in this case adds to the evidence of 
the self-affine nature of the height fluctuations. 
The third moment of the scale-dependent curvature confirmed that
the height fluctuations are non-Gaussian, while 
the contour-loop fractal dimension and size distribution
indicated self-affine scaling with 
$\alpha\approx 0.85$.

Experimental data often exhibit self-affine scaling up to a correlation 
length $\xi(t)$. We argued (in Sec.~\ref{subsec_perc}) that 
the loop exponents, for loops whose linear size exceeds 
the correlation length, are determined by exactly known percolation exponents. 
The crossover between the self-affine and percolative regime was visible
(with a consistent $\xi$ value) in every kind of measure
on the experimental data in Sec.~\ref{sec_exp} -- the same was
true for Gaussian random surfaces with an artificial length scale cutoff
(Sec.~\ref{subsec_gauss_finitecorr}).
The numerical values of the percolative exponents were confirmed
from simulations of Gaussian surfaces with a white-noise spatial
power spectrum. 

Our results (see Sec.~\ref{sec_gauss}) show that it is 
quite difficult to get correct results
from loop measurements when $\alpha$ is near to 1. 
The reason, we believe,  
is that the crossover to asymptotic behavior occurs at very large loops; the  
inferred $\alpha$ is thus smaller than the real one.
It has been observed \cite{yang} that even the height-height
correlation function tends to yield a too small value of $\alpha$ 
as compared to the Fourier power spectrum, even though the two 
measures have, in principle, the same information.

\subsection {Comparisons of roughness measures}

Roughness has often been analyzed based on a single 
number, the overall variance of the height over the entire system.
However, spatial correlations in height fluctuations are central
to the development of self-affine or other interesting morphologies.
Therefore, every form of roughness measure we discuss takes the
form of a spatial {\it spectrum}, i.e., one measures an entire function 
whose argument has dimensions of length (called $s$, $R$, $b$, or $1/q$).
The variation of the roughness measure with its argument is
related to the varying amount of interface fluctuations on the 
corresponding length scales. 

Some previous measures of the self-affine exponents were
reviewed in Ref.~\onlinecite{schmitt}. 
They systematically compared the different measures
using artificially constructed realizations of $h(x)$
(only in $1+1$ dimensions), and 
concluded that the single best
measure of $\alpha$ is the Fourier power spectrum, 
our \Eq{power}. 
(Oddly enough, Ref.~\onlinecite{schmitt} did
not include the height correlation function, our \Eq{struc}, 
in their selection of measures to compare.)

Another approach is to measure $h({\bf x})$ along a single
line in the $\bf x$ plane, corresponding to a line scan 
by the STM\cite {tong}.
This section through the surface
may then be analyzed as if it were a $1+1$-dimensional profile. 
Ref.~\onlinecite{tong} evaluated the variance over an interval
of length $L_0$, which should scale as $L_0^{2\alpha}$, and
applied this experimentally to the heteroepitaxy of 
CuCl on the (111) surface of $\rm CaF_2$. 

The most useful roughness measures have been discussed and 
critiqued in the sections related to them;
they fall into three categories and are
summarized here in Table~\ref{tab_measures}.

\subsubsection {Quadratic roughness measures}

The most familiar measures are {\it quadratic} 
of which the first three were summarized in Sec.~\ref{subsec_quadmeas}.
Besides three well-known quadratic measures, 
we include a fourth which has {\it not}
been previously applied: the variance of scale-dependent curvature, 
$\langle C_b({\bf x})^2\rangle$, which we introduced in \Eq{curv2} 
(Of course, the ensemble expectation cannot depend on $\bf x$.)
Its behavior is very similar to that of the height-difference correlation
$D_2({\bf r})$, so $\langle C_b^2 \rangle$
is of interest mainly for comparison with the higher moments
of $C_b({\bf x})$. In practical applications
the Fourier spectrum is probably the best of these. 

A key fact about the quadratic measures is that, given the complete 
function for any one of them,
one can compute the complete
function for any other one as a linear transform 
(convolution with some kernel) 
This property is not true for higher moments. 
Notice also that the quadratic measures are invariant under
$h({\bf x}) \to - h({\bf x})$ and so cannot possibly characterize the breaking
of up/down symmetry in the growth process.  
Nor can they identify deviations from Gaussianness, since 
one can produce a Gaussian ensemble (as in Sec.~\ref{sec_gauss_construct})
with any given Fourier spectrum.

\subsubsection {Non-quadratic roughness measures}
\label{non_quad}

Essentially all of these have been developed by
analogy with quadratic measures, simply replacing the second power
by a higher power. 
Our curvature-moment function seems to be the first generalization
of the height-difference function that captures the up/down asymmetry.

A simple  generalization of the $b$-box variance
is the $b$-box $q$-th moment, 
$\langle (h({\bf r})-\overline{h}_b)^q\rangle _b$. The $q=3$ moment 
characterizes the up/down asymmetry; when scaled by 
$\langle (h({\bf r})-\overline{h}_b)^2\rangle _b^{3/2}$ it defines a
scale-dependent, dimensionless skewness 
that measures the deviation from Gaussianness\cite{racz}.
This appears to be  a simple and attractive measure, but we
know of no applications to date; our curvature moment 
$\langle C_b^3 \rangle$ is similar in spirit, but probably
not linearly related. 

We evaluated the quartic curvature moment $\langle C_b ^4 \rangle$, 
but this data was less useful than our other measures: 
it does not reveal the non-Gaussian nature as strikingly as
$\langle C_b ^3 \rangle$ does. 
The dimensionless ratio 
$\langle C_b ^4 \rangle / \langle C_b ^2 \rangle ^2$ is 3 in the
Gaussian case, but may not differ very much in a non-Gaussian
ensemble. Furthermore, the roughness exponent was fitted 
less precisely from $\langle C_b^4 \rangle$ than from any other measure, 
probably due to the sensitivity of higher moments to rare events. 

The analysis in Ref.~\onlinecite{kleban} summarized in
Sec.~\ref{subsec_kleban}, 
appears to be the first
application of a scale-dependent roughness measure to characterize the
up/down asymmetry. However, we believe a
 local roughness measure such as the 
$b$-box skewness or (better) our mean cubed curvature gives a
more meaningful characterization.
In a sense, 
the smoothed-height skewness is the opposite of the local measures
since it includes the fluctuations from all length scales {\em larger}
than $b$,  while the local measures include the fluctuations from
scales comparable to or smaller than $b$; only the latter would be
expected to scale as $b^\alpha$. 

\subsubsection {Loop measures}

The other non-quadratic roughness measures, of course, 
are the loop measures defined in Sec.~\ref{sec_measures}. 
The length and connectedness of a loop, manifestly,
depend on the heights $h({\bf r})$ in a highly nonlinear fashion, and
one might expect the loop exponents to be independent of the
roughness exponent $\alpha$;  then the loop properties
might have distinguished between different universality classes 
of growth which happen to have 
similar $\alpha$ values.    From this viewpoint, it is disappointing that 
we in fact find the loop exponents are functions of $\alpha$
(Sec.~\ref{sec_scaling}). 
Thus for self-affine interfaces 
the loop measurements serve only as a check on other ways 
(quadratic and non-quadratic) of measuring $\alpha$. 
Furthermore, when the heights at large separations are
uncorrelated, implying the loops are percolation hulls 
(see Sec.~\ref{subsec_perc}), 
the loop plots show a weaker change of slope at this crossover than 
the Fourier spectrum does. 

It seems worthwhile nevertheless to compute loop measures. In a sense
they depend on higher order correlation functions of the heights: 
then the agreement between the $\alpha$ values extracted from loops
and from other measures is an additional, stringent test of self-affineness. 
Also we observe empirically 
that loop measures, and in particular the average loop length
as a function of loop radius, are very self averaging and measurement 
of $\alpha$ from them produces smaller errors than either the real-space 
or Fourier-space methods.
Finally, although the loop exponents
are the same for different universality classes with the same $\alpha$, 
we do expect universal coefficients to be different.

For computer generated height data
the loop-size-distribution is, perhaps,  the single most
valuable plot, because two different exponents can be obtained
from scaling plots such as \Fig{Fig:FSS_PS_04}.
This is not the case for experimental data where the system size 
is typically not a tunable parameter.

The loop-correlation function $G({\bf r})$ is most tedious to compute, and
since its exponent $2x_l$ is superuniversal it does not yield
an estimate of $\alpha$. Nevertheless $G({\bf r})$ is a useful check
on the self-affineness, since the superuniversal behavior fails in
other cases (e.g. beyond the correlation length, see \Eq{perc_results}).

\subsection{Future directions}

New experimental techniques  which provide
complete real-space images of the fluctuating quantity of 
interest (rather than system-wide averages, or
local measures probing the system at only a few points), 
are being developed in every physical science.  
Consequently, measures which usefully exploit
this wealth of information will gain in importance.   
In turn, the ability to measure new (and nonlinear) correlations
may inspire new theories that can predict the correlation behavior.

\subsubsection{Turbulence}

Fluid dynamics is a good example of the interplay just mentioned 
between theory and experiment: formerly two (sometimes more)
point correlations were measured by hot-wire probes, and the same
correlations were the objects of the Green's function method.
As full images become available of the velocity field, 
many new measures are attempted in order to capture more
of the available information.

Indeed, the measures introduced here 
might be adapted to the geometrical description of turbulence. 
The advection of passive tracers by turbulent flows seems 
a to be an especially  promising problem. 
There is already considerable interest in characterizing the
equal-time correlations through fractal measures of the contours
of (say) constant tracer concentration 
\cite{sreeniII,gollub,catrakis,tabeling}. 
(To maintain the analog of a surface's symmetry under global
shifts of the height, one should study the {\em logarithm} of
the concentration and use contours spaced equally on the 
logarithmic scale.)

Measurements of the fractal dimension of iso-concentration lines of a 
passive tracer advected by a magnetically driven, turbulent, two-dimensional 
flow were reported by Cardosa  {\em et al.} \cite{tabeling}.
They find $D_f=1.35(5)$, which, assuming the concentration field is 
self-affine, yields a roughness exponent $\alpha=3-2D_f=0.3(1)$. 
Indeed, $\alpha=0.30(3)$  was measured by the authors, by applying the 
$q=1$ multiaffine 
correlation measure (entry 6. in Table~\ref{tab_measures}). We therefore
infer that their measurements are consistent with a self-affine morphology
for the concentration field.  
Details of the complete  loop and curvature analysis of this data set will 
be reported in a separate paper\cite{JKGreg_unp}.  

\subsubsection {Other dimensions?}

In this connection, it is interesting to consider the generalizations
of  $h({\bf x})$ to spatial  dimensions of $\bf x$ other than 2.
In the $1+1$-dimensional case, there are no loops;
the probability of first return to a fixed height value~\cite{schmitt}
seems to be the closest analog to our loop
size distribution of Subsec.~\ref{Length-dist}
(if presented as a distribution of $R$ instead
of $s$) and also to our loop correlation function \Eq{loopcorr}.

With each higher dimensionality there is greater
richness in distinct geometrical measures that can be defined for
iso-surfaces.
For a hypersurface in 3+1 dimensions -- like the concentration function 
in three-dimensional passive tracer advection -- 
the level set may be multiply connected  and even knotted.
Nevertheless the size distribution exponent $\tau$, 
the fractal dimension $D_f$, and $x_l$ of the connectedness correlations, 
can be generalized directly. But we see much less reason to expect a 
super-universal connectedness correlation exponent in dimensions higher 
than  $d=2+1$.

\subsubsection {Multifractality and scaling relations}

Several mysteries remain about the scaling relations derived in
Sec.~\ref{sec_scaling}. Above all, there is not yet any rigorous
or analytic basis for our fundamental conjecture 
(\Eq{loopcorr-conj}) of a super-universal loop correlation that scales as
$1/r$ for all rough self-affine surfaces -- unaffected even by
quenched disorder that further roughens the interface~\cite{chen_disorder}.
A second open question is to check numerically the
correlation exponent $a$ in \Eq{Gsform} for an individual loop 
of fixed size; we did not evaluate it in any of our numerical
studies, but it should not be the same as the exponent $2x_l$ for 
the ensemble average over loops of all sizes. 
Finally, it is intriguing to ask what happens in 
a ``multiaffine'' system \cite{krug_PRL,DS-punyindu}.
Here different moments of the height variables have
different scaling exponents; which of these (if any) is
the one entering our formulas (\ref{results1}) and 
(\ref{results2}) for the loop exponents?

\acknowledgements

We  would like to acknowledge helpful discussions with M. Aizenman, 
M. Avellaneda, P. Constantin,  
D. Fisher, G. Huber, M. Isichenko, T. Spencer, C. Zeng, 
and particularly to thank J. Krim, M. Plischke and  Z. Csahok
for supplying STM data, and 
Michael Plischke for providing his code to simulate the single-step model.
We are grateful to M. B. Isichenko for bringing  Ref.~\onlinecite{avella}
to our attention, and to J. Gollub for alerting us to 
Ref.~\onlinecite{tabeling}.
C.L.H. was supported by NSF Grant No. DMR-9612304, 
D.G.S. by NSF Grant No. DMR-9632275 through the Cornell Materials
Science Center, and J.K. by 
NSF Grants No. DMS 93-04580, DMS 97-29992,  and
(during a portion of this work at Brown University) DMR-9357613.

\appendix
\section {Analytic derivation of $x_l(\alpha=0)$ for an exact soluble model}
\label{app_O2loop}

Our purpose here is to
support the conjecture (\ref{loopcorr-conj}) in Sec.~\ref{sec_loopexp}, 
by showing that $x_l=1/2$ in the case of a 
lattice model which can be mapped to an equilibrium-rough surface.
At long wavelengths height fluctuations are described by  
the well-known free energy 
  \be{eq-rough}
   F = ({\rm constant}) \int d^2{\bf x}  |\nabla h({\bf x})|^2
  \ee
which by equipartition implies Eq.~(\ref{hqvar}) with $\alpha=0$, 
so indeed the surface is self-affine.
This appendix only summarizes arguments made previously
in Refs.~\onlinecite{loop-PRL}, \onlinecite{jk-4col}, and
\onlinecite{jk_NPB}. 

Consider a  statistical model with  microscopic heights $z_{\bf j}$ 
defined on a triangular lattice $\{\bf j\}$, 
such that $z_{\bf j}$ changes by 0 or $\pm 1$ 
between nearest neighboring sites.
The partition function of the model is 
\be{z-part} 
Z = \sum_{\{z\}} \prod_{<{\bf j}, {\bf k}>} \: w(z_{\bf j} - z_{\bf k})
\ee
where $w(0)=1$ and $w(\pm1)=K$; the sum goes over all microscopic height 
configurations unrelated by a global height shift.

A contour-loop configuration $\gamma'$ is specified by drawing  
closed (periodic boundary conditions ensure that all contour lines are 
closed), {\em oriented},  non-intersecting loops along the bonds of the 
dual honeycomb lattice, which seperate sites that differ in height by $\pm 1$
(the sign determines the loops orientation).
In terms of the loops, the partition function is 
\be{cl-part} 
Z  = \sum_{\gamma'} K^{N_b}
\ee
where $K$ is the fugacity of an occupied bond (i.e.~one covered by a 
loop), and $N_b$ is the number of occupied bonds in $\gamma'$. 

This model is equivalent to the $O(2)$ loop model introduced by 
Nienhuis\cite{nien_rev}. 
This is seen by  rewriting the partition function in terms of 
non-oriented loop configurations $\gamma$,
\be{o2-part}
Z = \sum_{\gamma} K^{N_b} 2^{N_l}
\ee
where $N_l$ is the number of loops in $\gamma$ (which is the same 
as the number of loops in $\gamma'$), and the $2$ appears as a result of 
summing  over the two possible orientations for each loop in $\gamma'$. 

By mapping the $O(2)$ loop model to the 4-state ferromagnetic 
Potts model on the triangular
lattice, Nienhuis showed that for $K_c=\sqrt{2}$ the loop model is
critical.
Using the Coulomb-gas picture of correlations~\cite{nien_rev} this
implies that $\{z_{\bf j}\}$ are rough; since
the ``background charge'' is zero\cite{nien_rev}, 
it is plausible that (at $K_c$) 
the height model is equilibrium-rough, i.e. the field
$h({\bf x})$ obtained by coarse-graining
$z_{\bf j}$ satisfies (\ref{eq-rough}).
Furthermore
the contour-loop correlation function can be identified in the $O(2)$ model 
with the so-called energy-energy correlation function which at 
$K=K_c$ decays as a power law with the known exponent 
\be{saleurx1}
x_l=1/2  \ , 
\ee
as was to be shown.

Since we view the $O(2)$ loop model simply as one of many possible 
lattice disretization of a 
logarithmically rough ($\alpha=0$) self-affine surface, 
then the exponent $x_l = 1/2$ should  necessarily  appear in other
lattice models that map to rough surfaces. 
Indeed the same 
value of this exponent follows also from the exact solution of the
$O(2)$ loop model on the square lattice\cite{On_sq}, 
and the $n=2$ fully packed loop model on the honeycomb lattice\cite{batch94}.

\section {Percolation scaling of contours for uncorrelated heights}
\label{app_perc}

This appendix derives the scaling 
behavior of the loop ensemble when the 
random heights $h({\bf x})$ have a finite variance and
(beyond a correlation length $\xi(t)$) are {\em uncorrelated};
this describes early stages of growth, as in 
Sec.~\ref{subsec_perc}.

To model the contour loops at length scales greater than $\xi(t)$, 
first coarse-grain the system into boxes of side $\xi(t)$.
The average height $h$ in each box
is an independent random variable parametrized by
$p(h')$, the probability  that $h<h'$. 

Defining all the boxes with $h<h'$ as ``filled'' simply reproduces the 
(uncorrelated) percolation clusters for occupancy $p(h')$. 
Then every contour of constant $h'$ is
simply the perimeter of such a cluster. 
This mapping is well-known from Ref.~\onlinecite{trugman}
and is widely applied in the theory of the 
quantum Hall effect\cite{isich,trugman}.

The percolation clusters -- as well as their perimeters --
are self-similar only at 
$p(h)=p_c$, the percolation threshold; we will first discuss their
(known) loop exponents. The behavior when $p\neq p_c$ can easily
be derived from well-known percolation scaling relations.
The final step will be to integrate these results over $p$, since
the loop ensemble we simulate actually corresponds to the union of 
perimeter ensembles for all $p$.

\subsection{Contour loops and critical percolation}

Fixing $p=p_c$ for a moment,  the perimeter loop ensemble may be characterized
by exponents $D_h$  and $\tau_h$, with definitions analogous to 
(\ref{dimdef}) and (\ref{taudef}) for $D_f$ and $\tau$.
(The subscript ``$h$'' stands for ``hull'' as the perimeter is often called.)
The fractal dimension 
  \begin{equation}
     D_{h}=7/4
  \end{equation}
is known exactly\cite{saleur}. 

The perimeter loops  for percolation at $p_c$
also satisfy a hyperscaling relation analogous to (\ref{scaling2}), 
with $\alpha$ replaced by zero. 
That is, the largest cluster (or perimeter) diameter 
inside a box of side $l$ is least $\sim l$. 
From this follows a relation for $\tau_{h}$ 
  \begin{equation}
     \tau_{h} = 1 + 2/D_{h} = 15/7 \ . 
  \end{equation}

When $p \neq p_c$, the cluster (and perimeter) ensemble scaling 
is cut off at the percolation correlation length $\xi_p(p)$, which 
diverges near $p_c$ as 
  \begin{equation}
    \xi_p(p) \sim |p-p_c|^{-\nu_p} \ ,
     \label{xip}
  \end{equation}
where $\nu_p$ is the usual percolation correlation exponent, 
and $\nu_p =  4/3$ is known 
exactly~\cite{dennijs-perc,saleur}.
In this case, the loop length distribution is
  \begin{equation}
     P_h(s;p) = s^{-(\tau_{h}-1)} f_h(s/\xi_p(p)^{D_{h}}) 
     \label{Ppsp}
  \end{equation}
where $f_h()$ is a  scaling function, which falls off exponentially fast
for loops of radius greater than $\xi_p(p)$. 

\subsection {Union of all percolation contours}
\label{perc-union}

In the percolation regime, evidently, the statistical properties
of the contours of a particular 
level set  depend on the chosen level $h$. 
(This was impossible in the self-affine regime, since
in that case the fluctuations of $h$ were unbounded.)
But we have previously studied the {\it union} of all contours
with different $h$, 
corresponding to all values of $p(h)$  from 0 to 1. 
That is, indeed, the ensemble sampled by our computer codes
(see Sec.~\ref{sec_loopmeas}). 
We will now derive 
the exponents $\tau_{p}$ and $x_{l,p}$ of {\it this} ensemble, 
defined analogously to $\tau$ and $x_l$ in eqs.
(\ref{taudef}) and (\ref{loopcorr}). 

Most of the loops at a large length scale $R$ come from levels sets at 
height $h$ with $\xi(p(h))> R$,  rather than from the exponential tails
of the distribution (\ref{Ppsp}) for the other $h$ values. 
Thus these obey the percolation scaling and 
all have fractal dimension $D_{f,p} \equiv D_{h}$. 
This behavior is illustrated in Fig.~\ref{fig_perccross}(c). 

Now, in the percolation regime, $P(s)$ as defined in Sec.~\ref{Length-dist}, 
is just proportional to the integral of $P_h(s;p(h))$ over h. 
A  weighting factor $|dp/dh|$ should be included
as the contours are equally spaced, and
$p(h)$ is normalized to unity. 
Since the large contours come from $p\approx p_c$, 
only that part of the distribution matters.
Inserting (\ref{xip}) into (\ref{Ppsp}), one obtains 
  \begin{equation}
     P(s) \sim \int d p ~ s^{-(\tau_{h} - 1)} 
                    f_{h}(({\rm const}) s |p-p_c|^{D_{h}\nu_p} )
     \label{Pssum}
     \end{equation}
  hence   $P(s) \sim s^{-(\tau_{p}-1)}$   with 
  \begin{equation}
      \tau_{p} = \tau_{h} + (D_{h} \nu_p)^{-1} = 18/7 \ .
       \label{taulp}
  \end{equation}

Finally, given (\ref{taulp})
the simplest route to the loop (connectedness) correlation exponent 
is to use the exponent relation (\ref{scaling1});
this gives
  \begin{equation}
     2 x_{l,p} = 4-2 D_h + \nu_p^{-1}  = 5/4 \ .
     \label{xlp}
  \end{equation}
Eq.~(\ref{xlp}) could alternately be reached by first noting
that the corresponding exponent is $1/2$ for the percolation
hull ensemble at $p_c$, and then averaging the loop connectedness
correlation function analogous to (\ref{Pssum}). 

\section{Loop finding algorithm}
\label{app_loopalg}

Given a square lattice ${\cal L}$ on which the heights $h$ are defined  
and a point ${\bf x}_0$ on the dual lattice ${\cal L}^\ast$,  
the task of the loop finding 
algorithm is to construct a contour loop of the surface  
which passes through the point ${\bf x}_0$. 

The contour is a walk along the bonds of  ${\cal L}^\ast$ that cuts 
those bonds of ${\cal L}$
that have vertices with heights lying above and below the 
contour height, \Fig{fig_walks}. To implement this idea we first 
define the level height $h_{\rm lev}$ which is the average of the four heights 
around the  plaquette centered at ${\bf x}_0$. 
Second, we  assign to all the sites of ${\cal L}$ $+$ or $-$ signs 
according to whether
they are above or below the chosen level $h_{\rm lev}$. Now, starting 
from ${\bf x}_0$ we form the contour loop by drawing links on the dual 
lattice which cross the bonds of ${\cal L}$ connecting
$+$ and $-$ sites. This is repeated  until the walk  returns
to the starting point ${\bf x}_0$; the finite extent of the lattice ${\cal L}$
is dealt with by implementing periodic boundary conditions. 

Special care must be taken whenever a ``saddle-point'' plaquette is 
reached, that is one  where the sites of the lattice   
are assigned  $+-+-$ signs  cyclically around the plaquette.
In this case four links meet 
at the point in the center and we must resolve the connectivity there
by an additional rule, so as to convert this pattern into two
90-degree turns that are not quite touching.
One natural condition on the rule is that it should 
be reversible, that is  one should find the 
same loop whether one starts traversing it clockwise or counterclockwise.
A second condition is that it ought to be invariant under reflecting
all heights by $h({\bf x}) \to -h({\bf x})$.  
A  physically sensible rule which satisfies both conditions 
makes use of the average height $h_{\rm plaq}$ of the four 
heights around the saddle-point plaquette. 
If $h_{\rm plaq} < h_{\rm lev}$, we view
the center of the plaquette as being lower than the level of the 
contour loop and the connectivity is resolved by having the $+$ sites 
{\em inside} the 90-degree turns. In the opposite case,  
$h_{\rm plaq} >  h_{\rm lev}$, 
the $+$ sites lie {\em outside} the 90-degree turns; see \Fig{fig_walks}.
(The agreement of loop data from
the single-step model with parameters $p_+$ and with $1-p_+$, 
as explained in Sec.~\ref{sec_SSM}
was a valuable check of the up-down symmetry of our loop-finding algorithm.)

Once a contour loop through ${\bf x}_0$ has been constructed its length and
radius are recorded, assuming that the loop is topologically trivial. (Due 
to periodic boundary conditions loops with non-zero winding numbers are
possible and these we discard.) 
The contour loop length $s$ is equal to the number of steps made during the 
loop construction, 
while the radius $R$ is the size of the largest square which covers the loop, 
i.e., the maximum displacement in the $x$ or $y$ direction. 
Every topologically trivial loop also 
contributes to the correlation function $G(r)$; for every point on the loop
that is a distance $r\in[i,i+1)$ ($i$ is an integer) away from the starting 
point ${\bf x}_0$, the array element $g(i)$ is increased by one.   
We {\it define} $G(r)$ for our simulations as $g(r)/2\pi r$, 
which asymptotically is normalized the same as $G(r)$ defined in
Sec.~\ref{Length-dist}.

So far we have assumed that the height variables are real and the 
condition $h_{\rm plaq}=h_{\rm lev}$ is almost never fulfilled. This 
is not the case for interfaces which arise from discrete growth 
simulations like the one presented in section~\ref{sec_SSM}
where the height variable takes on integer values.
In this case the resolution of the connectivity should be completely
random  but we must ensure 
that we use the same choice if the loop returns to the same plaquette. 
The simplest way to do this, which is what we have implemented, is to  
take the original integer heights and ``dither'' them -- add
small amounts of random, uncorrelated gaussian noise to all 
$h({\bf x})$. This will
also solve the problem of choosing $h_{\rm lev}$; it will be
non-generic for any two heights to precisely coincide, although
this will happen occasionally as the price of roundoff error. When this 
does happen we start over by choosing a new initial site ${\bf x}_0$.

\newpage

\begin{table}
\caption{Geometrical exponents $x_l$, $D_f$, and $\tau$ for 
loops on Gaussian surfaces with various roughness exponents $\alpha$.
Columns marked ``direct'' are from direct fits 
to a power law of the data from system size $L=512$,
inferring $2x_l$, $D_f$, and $\tau-2$ from plots such as
Figs.~\ref{Fig:Gr_512}, \ref{Fig:Sr_512}, and \ref{Fig:PS_512}.
Columns marked ``FSS'' were fitted to finite-size scaling plots like
Figs.~\ref{Fig:FSS_Gr_04}(b) and \ref{Fig:FSS_PS_04}(b).
According to our conjecture,  the ``theory''
value of $2x_l$ is 1, independent of $\alpha$, 
and this is supported by the measurements here.
Notice a slight systematic deviation of the ``direct'' exponents 
from theory when $\alpha>0.5$, which we attribute to more severe 
finite size effects in those cases. 
The ``theory'' formulas for $D_f$ and $\tau-2$ are in \Eq{Df-alpha}.}
\label{tab_gauss}
\begin{tabular}{|c|cc|ccc|ccc|} 
$\alpha$ & \multicolumn{2}{c|}{$2x_l$} & 
          \multicolumn{3}{c|}{$D_f$} & \multicolumn{3}{c|}{$\tau-2$} 
           \\ \hline
        & direct & FSS &  direct & FSS & theory & direct & FSS & theory
           \\ \hline
0.0 &  1.07(2) &  1.02(2) & 1.48(1) &1.50(2) &   1.5 &  
        0.35(2) &  0.33(1) & $0.333\ldots$  \\
0.2 &  1.04(1) & 0.97(2) &  1.39(1) & 1.41(2) & 1.4 &  
        0.30(1) & 0.28(1) &  $0.286\ldots$  \\
0.4 &  1.01(1) &  0.98(2)&  1.31(2) & 1.32(3) & 1.3 &  
        0.24(1) & 0.225(5)&  $0.231\ldots$  \\
0.6 &   1.00(1) & 0.97(2) & 1.23(3) & 1.19(3)&  1.2 &  
        0.18(1) &  0.165(5) & $0.166\ldots$  \\
0.8 &   0.97(2) & 0.97(2) & 1.15(1) & 1.11(2)&  1.1 &  
      0.12(1)   & 0.11(2) &  $0.090\ldots$  \\
1.0 &   0.95(1) & 0.96(2) & 1.06(2) & 1.04(3) &  1.0 &  
      0.08(2)  & 0.02(2) &  $0.000\ldots$  \\ 
\end{tabular}
\end{table}

\begin{table}
\caption{Results of fits to roughness 
measures applied  to surfaces generated by the single-step model. 
The exponent values $\alpha$ in the left two columns
were derived in two direct ways, from the data plotted in Figures
\ref{Fig:SSM_C}(a) and \ref{Fig:SSM_hq}. 
The direct-fit results (``direct'') for $D_f$ and $\tau-2$ 
used only the $L=128$ data 
(shown in Figures \ref{Fig:SSM_SR} and \ref{Fig:SSM_PS});
the finite-size scaling results (``FSS'') were obtained using system 
sizes $L=32,64,128$.
The subheadings ``$\alpha$'' under
$D_f$ and $\tau-2$ are estimates of the roughness exponent
obtained from the FSS results by inverting \Eq{Df-alpha}.}
\label{tab_SSM}
\begin{tabular}{ c | c c | c c c | c c c }
$p_+$ & 
\multicolumn{2}{c |}{$\alpha$} & \multicolumn{3}{c |}{$D_f$}   &  
               \multicolumn{3}{c }{$\tau-2$}  \\   \hline
      & $\langle C_b ^2 \rangle$ &   $\langle |{\tilde h}({\bf q})|^2\rangle$ &
                  direct &  FSS & $\alpha$   &
                  direct &  FSS  & $\alpha$ \\   \hline
0.1  & 0.33(2) & 0.35(1) & 
        1.38(1) & 1.35(2) & 0.30(4) & 0.30(1) & 0.24(1) & 0.37(3) \\ 
0.3  & 0.19(4) & 0.09 (2) &
        1.47(1) & 1.46(2) & 0.08(4) & 0.38(1) & 0.35(1) & $-0.08(5)$\\
0.5  & 0.135(2) & 0.08 (1) &
        1.51(2) & 1.50(2) & 0.00(4) & 0.40(2) & 0.36(2) & $-0.13(5)$\\ 
\end{tabular}
\end{table}

\begin{table}
\caption{Roughness measures}
\label{tab_measures}
\begin{tabular}{ll}
Quantity  & Description \\
\tableline
\multicolumn{2}{l}{Quadratic measures} \\
\tableline
  1. $\langle (h({\bf x})-\overline{h}_b)^2\rangle _b$  &
    variance in $b \times b$ patch \\
  2. $D_2({\bf r}) = \langle |h({\bf r})-h(0)|^2\rangle$   &
    height correlation   \\
  3. $\langle |\tilde{h}({\bf q})|^2 \rangle $ &
    Fourier spectrum  \\
  4. $\langle |C_b ({\bf x})|^2 \rangle $ &
    ``$b$-dependent curvature'' variance  \\
\tableline
\multicolumn{2}{l}{Cubic and other non-quadratic measures} \\
\tableline
  5. $\langle (h({\bf r})-\overline{h}_b)^3\rangle _b$  &
     skew moment in $b \times b$ patch \\
  6. $\langle |(h({\bf r})-h({0})|^q \rangle$ &
     $q$-multiaffine height correlation \\
  7. $\langle (H_b({\bf r})-\overline{h})^3 \rangle$ &
     skew moment of $b$-smoothed height \\
  8. $\langle [C_b({\bf x})]^3 \rangle $ &
    ``curvature'' skew moment \\
  9. $\langle [C_b({\bf x})]^4 \rangle $ &
    ``curvature'' quartic moment \\
\tableline
\multicolumn{2}{l}{Loop measures} \\
\tableline
 10. $\langle s \rangle _R$ &
      average loop length, given radius $R$ \\
 11. $P_>(s)$  &
      Prob (loop through $\bf x$ is longer than $s$) \\
 12. $G({\bf r})$ &
      loop connectedness correlation function  \\
\end{tabular}
\end{table}

\begin{centering}

\begin{figure}
\begin{minipage}{8.66cm}
\epsfxsize=8.66cm \epsfbox{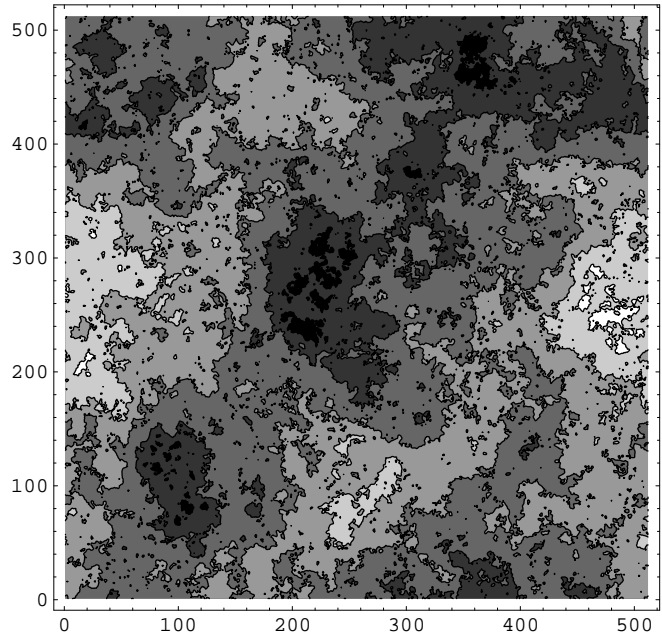}
\caption
{Contour plot of a $\alpha= 0.4$ random Gaussian surface; system 
size L=512.}
\label{fig_loops}
\end{minipage}
\end{figure}

\begin{figure}
  \begin{minipage}{8.66cm}
   \epsfxsize=8.66cm \epsfbox{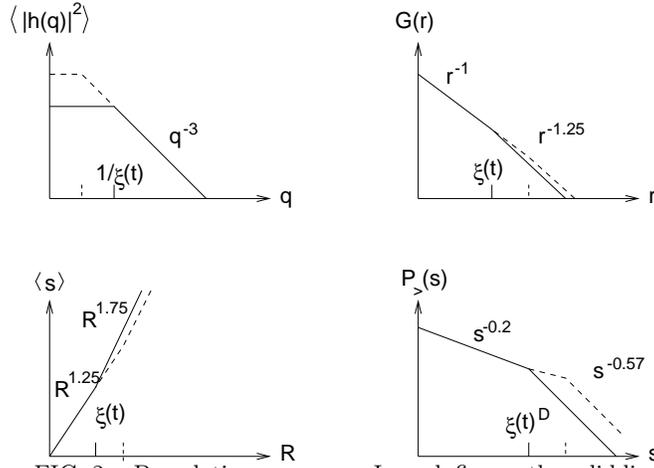}
   \caption{
Percolation crossover. 
In each figure, the solid line represents the function at a certain
time $t$, and the dashed line represented the same function at a 
later time when $\xi(t)$ has increased. 
The exponents shown in the figures are for $\alpha=1/2$, but the
qualitative behavior is the same so long as $0 < \alpha<1$. 
Each graph shows a crossover to percolation exponents
at a ``knee'' which corresponds to a length scale $\sim \xi(t)$:  
(a) Fourier spectrum $\langle |\tilde{h}(\q )|^2\rangle$,
(b) loop correlation function $G(r)$, 
(c) average loop length $<s>(R)$, and  
(d) cumulative distribution $P_>(s)$ of loop lengths (through a given point).
}
\label{fig_perccross}
  \end{minipage}
\end{figure}

\begin{figure}
  \begin{minipage}{8.66cm}
   \epsfxsize=8.66cm \epsfbox{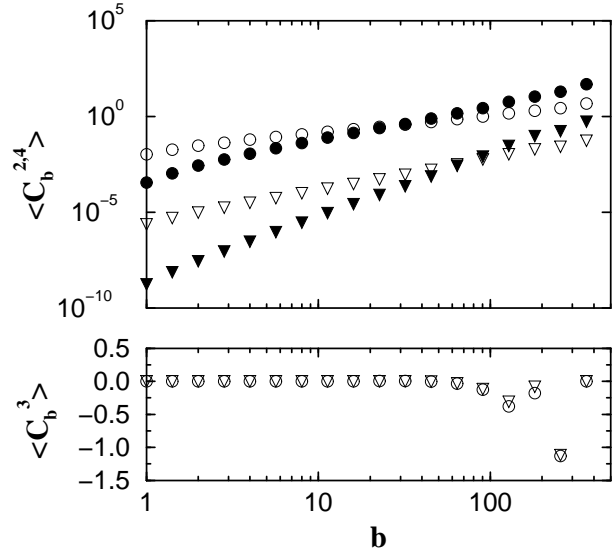}
   \caption{
Scale-dependent curvature moments from Gaussian 
random surfaces with roughness exponents $\alpha=0.4$ (circles) and 
$0.8$ (triangles).  
The third moments (lower plot) are zero confirming the
up/down symmetry. The upper plot shows
the second moments (open symbols) and fourth moments (filled symbols).}
\label{fig:curv_gauss}
  \end{minipage}
\end{figure}

\begin{figure}
  \begin{minipage}{8.66cm}
   \epsfxsize=8.66cm \epsfbox{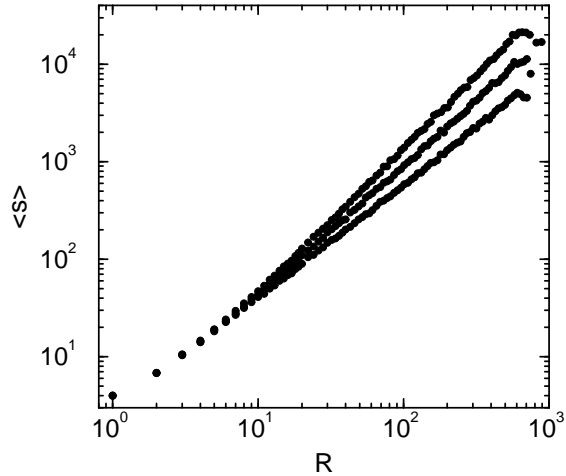}
   \caption{
Average loop length $\langle s \rangle$
as a function of loop radius $R$, for random Gaussian 
surfaces with $\alpha=0, 0.4, 0.8$ (from top to bottom); system 
size $L=512$, and 
$10^4$ loops were collected. 
The ``direct'' $D_f$ data in Table~\ref{tab_gauss}
are obtained by linear least-squares fits to such plots 
in the scaling regime, which is roughly $10<R<100$.
           }
   \label{Fig:Sr_512}
  \end{minipage}
\end{figure}

\begin{figure}
  \begin{minipage}{8.66cm}
   \epsfxsize=8.66cm \epsfbox{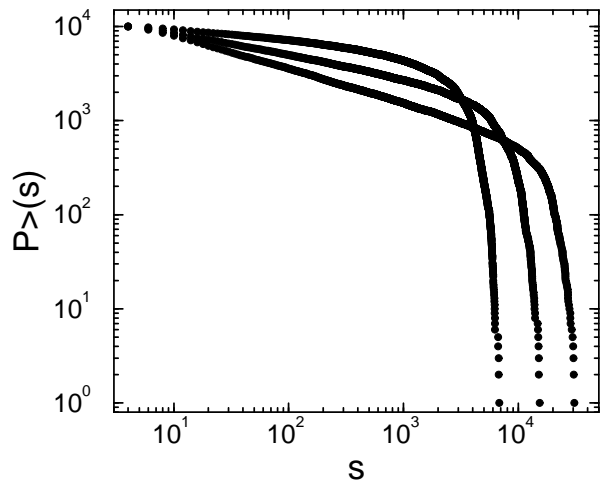}
   \caption{
Cumulative number of loops whose length is bigger than $s$ for random Gaussian 
surfaces with $\alpha=0, 0.4, 0.8$ (from top to bottom); system 
size $L=512$.
Here and in all other plots of $P_>(s)$, 
raw data is binned in intervals of form $(s,1.1s)$. 
           }
   \label{Fig:PS_512}
  \end{minipage}
\end{figure}

\begin{figure}
  \begin{minipage}{8.66cm}
   \epsfxsize=8.66cm \epsfbox{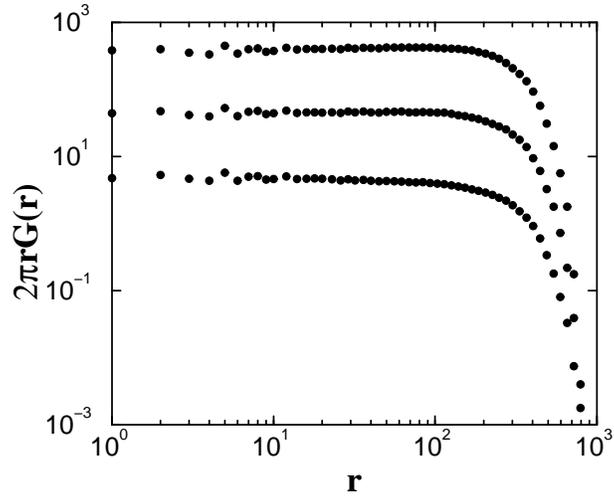}
   \caption{
Loop correlation function for random Gaussian 
surfaces with $\alpha=0, 0.4, 0.8$ (from bottom to top); system 
size $L=512$.  
In this and all such plots, raw data is binned logarithmically
in intervals of form $(r,1.1r)$. 
The latter two graphs are offset vertically by factors of 10 for clarity;
they are virtually identical except for a ``knee'' at slightly
different $r$ values.} 
   \label{Fig:Gr_512}
  \end{minipage}
\end{figure}

\begin{figure}
  \begin{minipage}{8.66cm}
   \epsfxsize=8.66cm \epsfbox{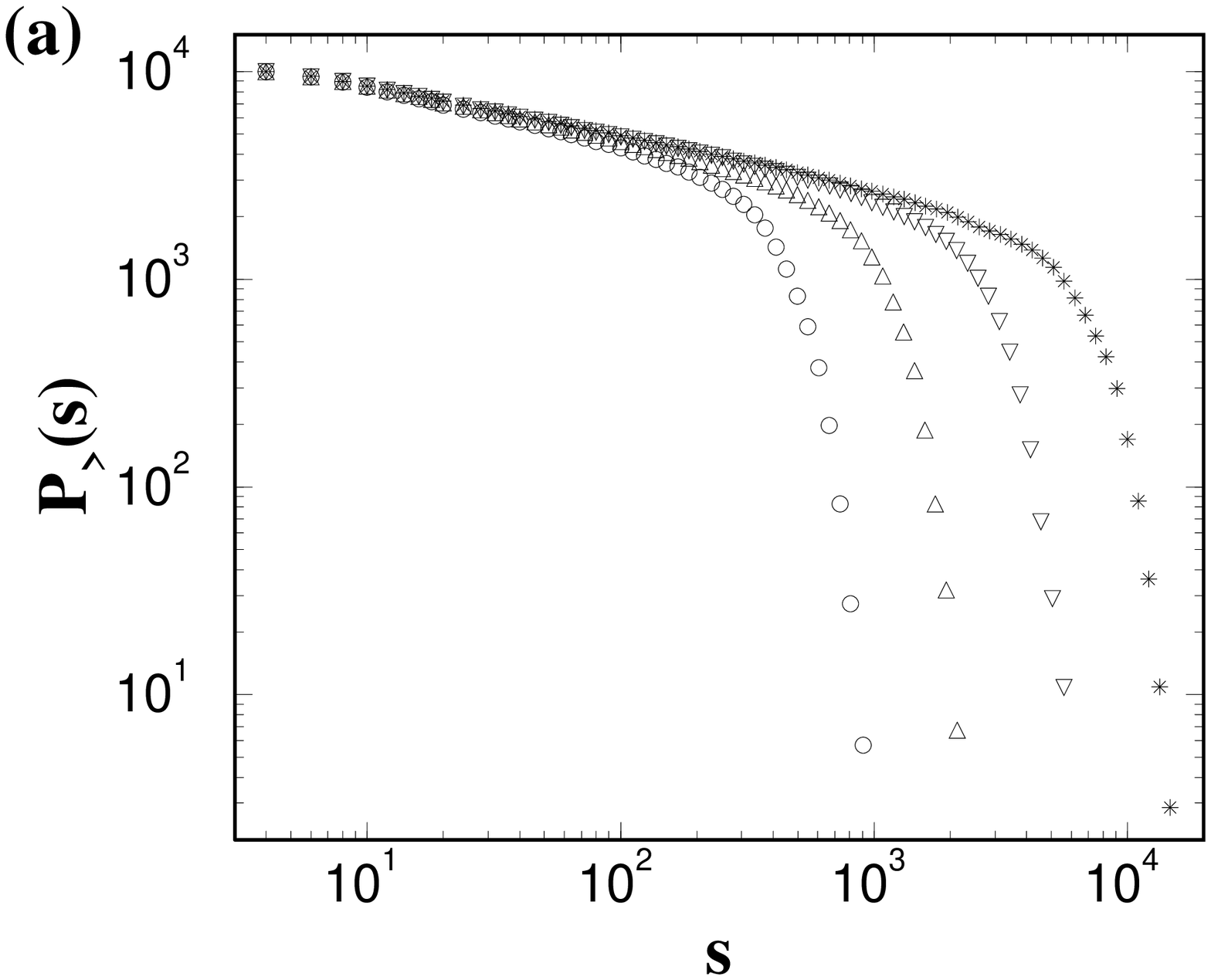}
   \epsfxsize=8.66cm \epsfbox{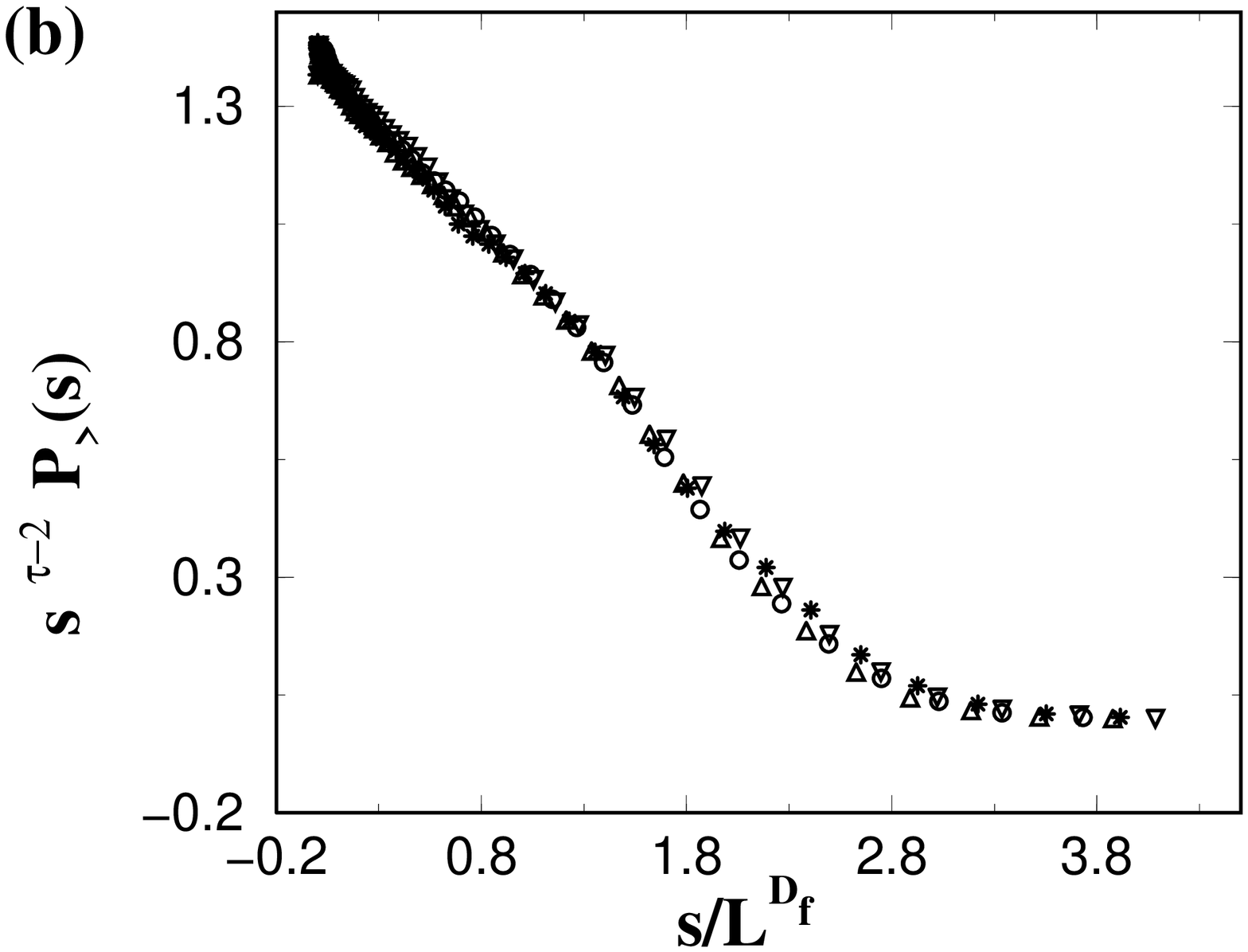}
   \caption{
Cumulative loop-size distribution, for random Gaussian surfaces with
$\alpha=0.4$. 
(a) Data for system sizes $L=64$ ($\bigcirc$), $L=128$ ($\bigtriangleup$)
$L=256$ ($\bigtriangledown$), and $L=512$ ($\ast$).
(b) Collapse of this data in a finite-size scaling plot with
$\tau-2=0.225$ and $D_f=1.32$. 
           }
   \label{Fig:FSS_PS_04}
  \end{minipage}
\end{figure}

\begin{figure}
  \begin{minipage}{8.66cm}
   \epsfxsize=8.6cm \epsfbox{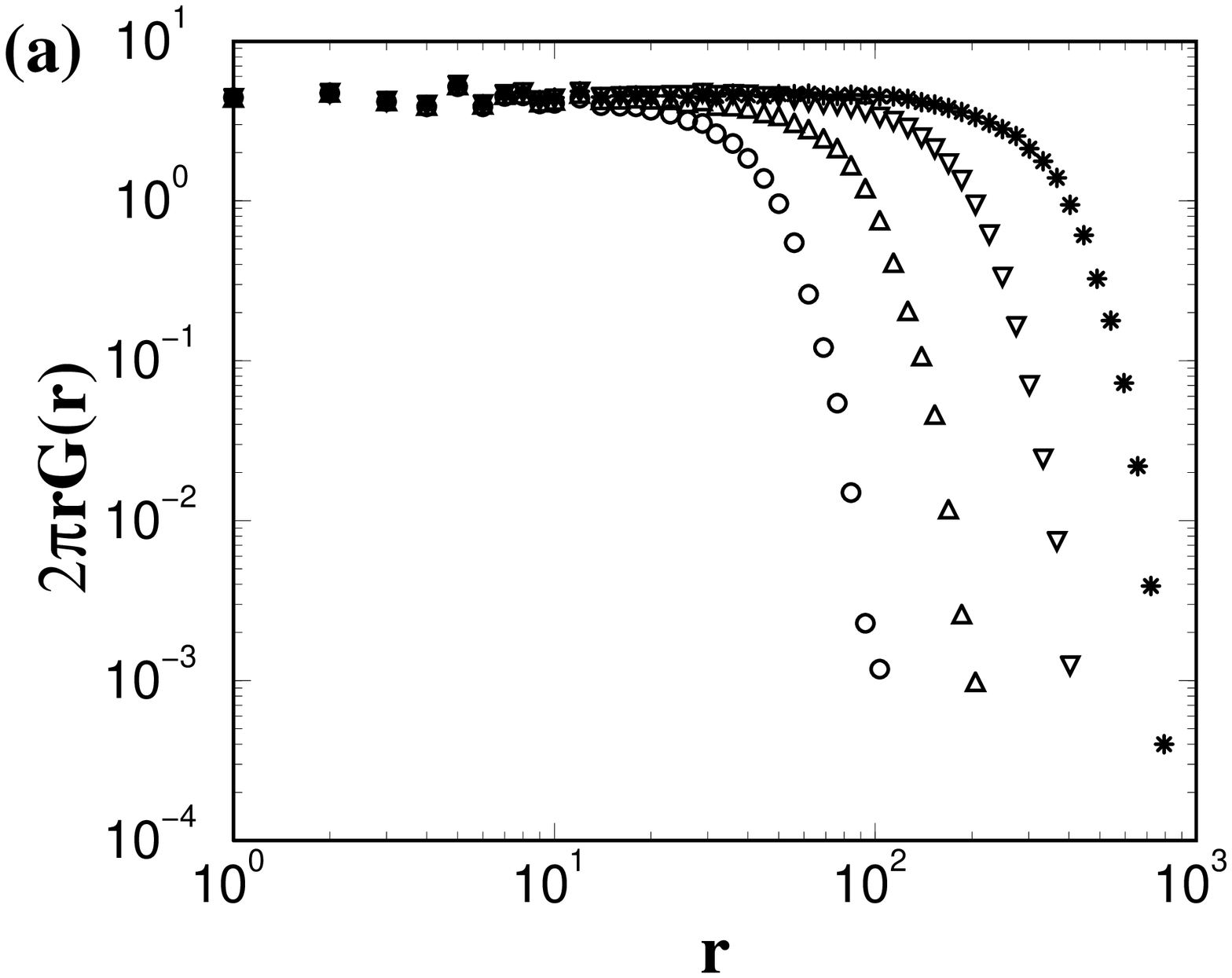}
   \epsfxsize=8.6cm \epsfbox{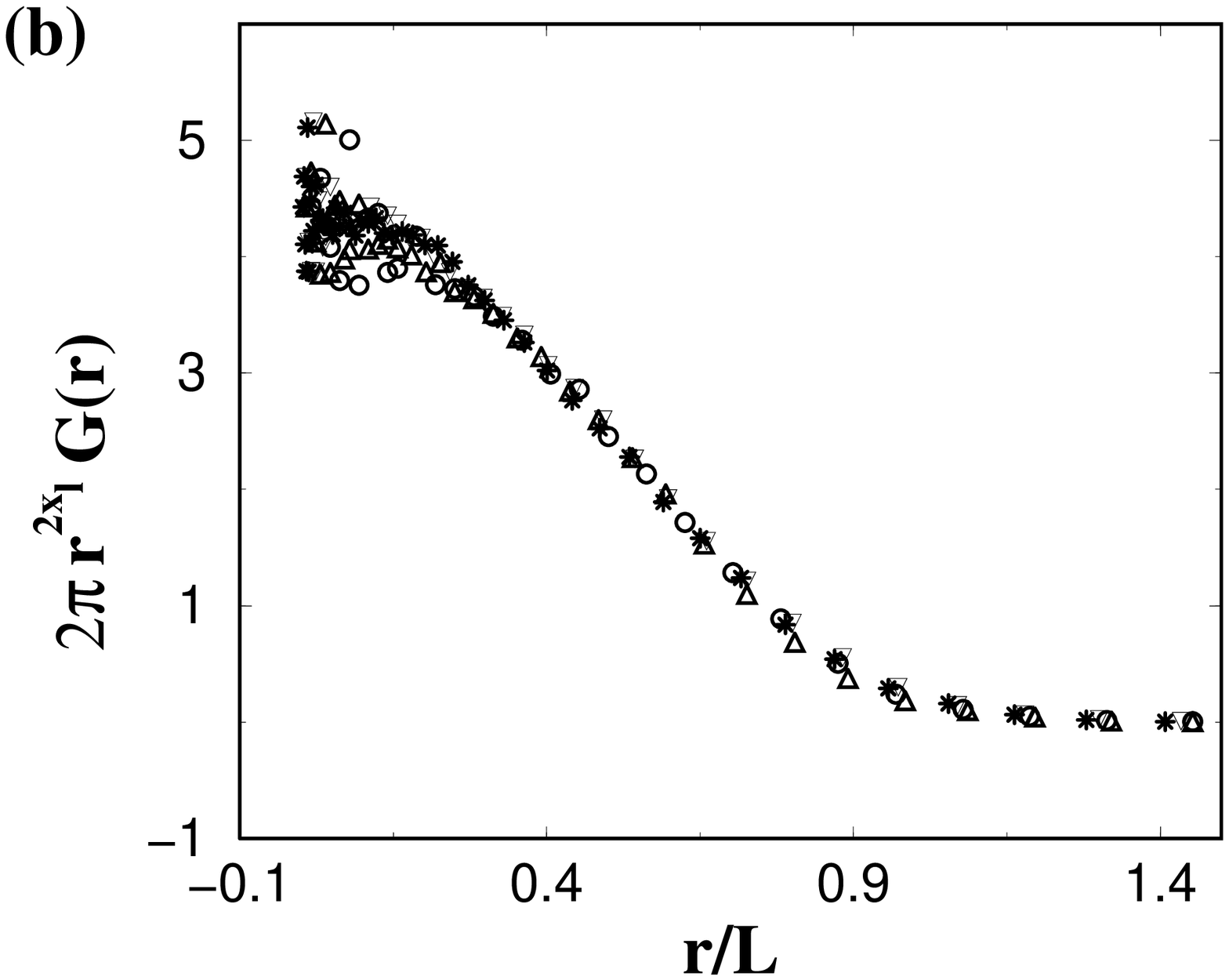}
   \caption{
System-size dependence of the loop correlation function $G(r)$
for Gaussian random surfaces with $\alpha=0.4$. 
(a) Data for sizes $L=64$ ($\bigcirc$), $L=128$ ($\bigtriangledown$)
$L=256$ ($\bigtriangledown$), and $L=512$ ($\ast$).
(b) Data collapse of this data with $2x_l=1.02$. 
           }
   \label{Fig:FSS_Gr_04}
  \end{minipage}
\end{figure}

\begin{figure}
 \begin{minipage}{8.66cm}
 \epsfxsize=8.66cm \epsfbox{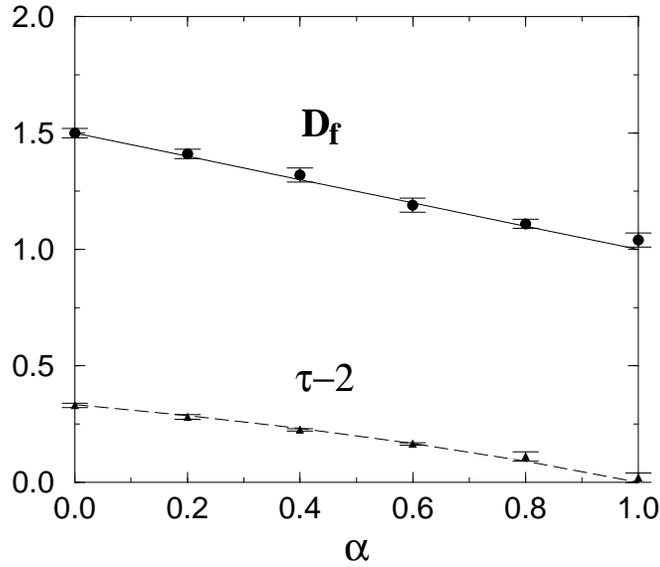}
\caption
{Fractal dimension $D_f$ and length distribution exponent $\tau-2$, 
as functions of the roughness exponent $\alpha$ of a
random Gaussian surface, obtained from finite-size scaling fits
(see Table~\ref{tab_gauss}).
The solid and dashed lines corresponds to the formulas 
in \Eq{Df-alpha}.
}
\label{Dfzeta}
  \end{minipage}
\end{figure} 

\begin{figure}
  \begin{minipage}{8.66cm}
   \epsfxsize=8.66cm \epsfbox{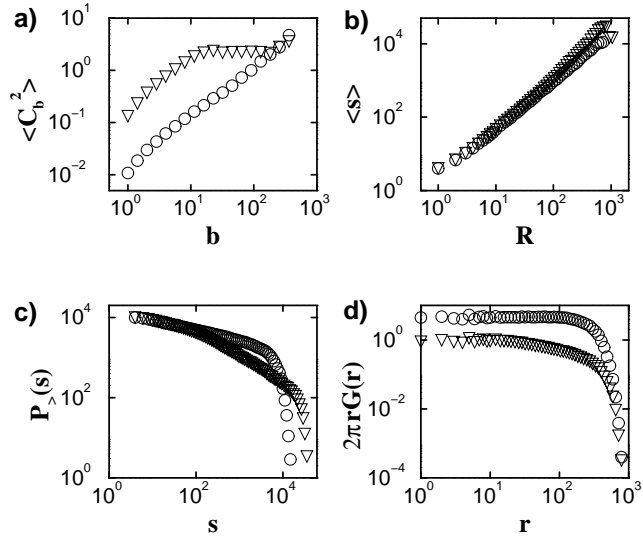}
   \caption{Finite correlation length effects for Gaussian random 
surfaces with $\alpha=0.4$, and a crossover to white noise for 
wavevectors smaller than $\pi/\xi_q$, where $\xi_q=16$; $L=512$ is the 
system size. Circles are used for data sets with  no cutoff which are  
included here for comparison with the cutoff data (triangles).  
a) Squared curvature function -- note the knee at $b\approx 15$.
b) Average loop length as function of radius -- knee at $R\approx 20$.
c) Cumulative distribution of loop sizes -- knee at $s\approx 100$.
d) Loop correlation function -- knee at $r\approx 20$.} 
\label{fig:gauss04_cut2}
  \end{minipage}
\end{figure}

\begin{figure}
  \begin{minipage}{8.66cm}
   \epsfxsize=8.66cm \epsfbox{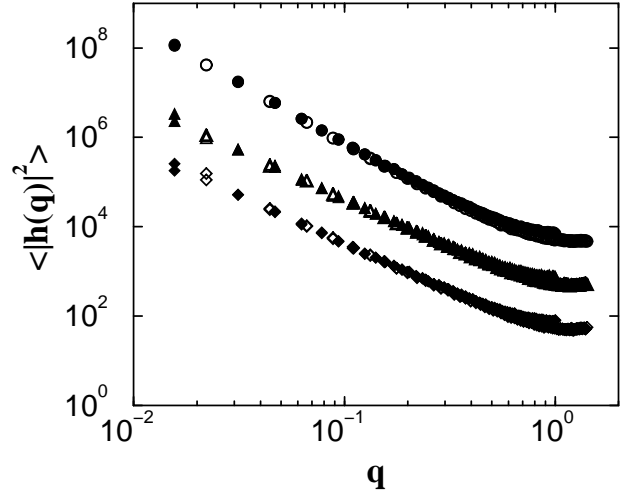}
   \caption{
Power spectrum of the height in the SSM model along the $[1,0]$ (filled
symbols) and the $[1,1]$ direction (open symbols) in reciprocal space. 
The data for $p=0.3$  (triangles) and for $p_+=0.5$ (diamonds) 
has been shifted with respect to 
the $p_+=0.1$ data (circles) by factors of $0.1$ and $0.01$ respectively
(for clarity). Note that the power spectrum is isotropic in Fourier 
space for small values of $|{\bf q}|$. }  
   \label{Fig:SSM_hq}
  \end{minipage}
\end{figure}

\begin{figure}
  \begin{minipage}{8.66cm}
   \epsfxsize=8.66cm \epsfbox{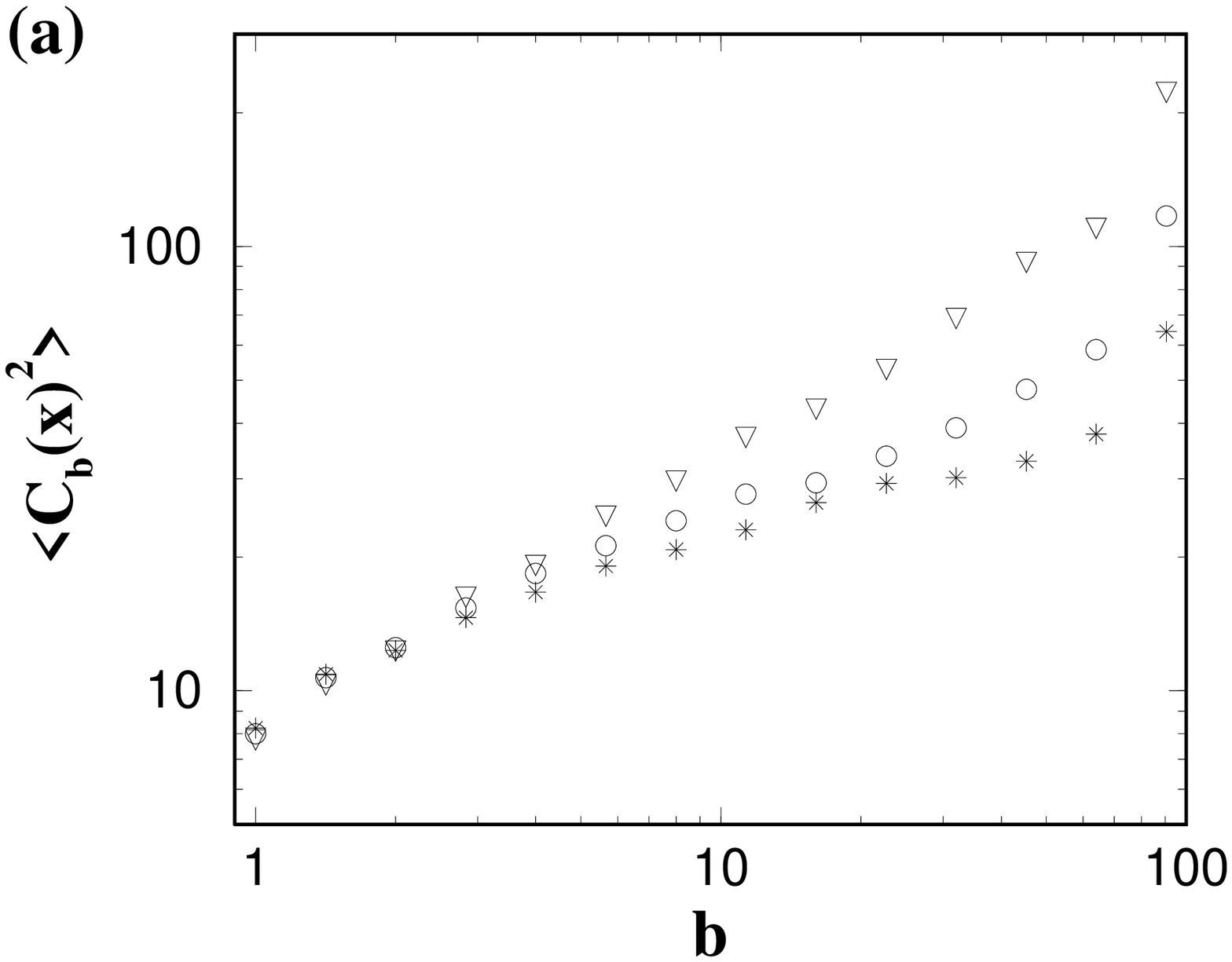}
   \epsfxsize=8.66cm \epsfbox{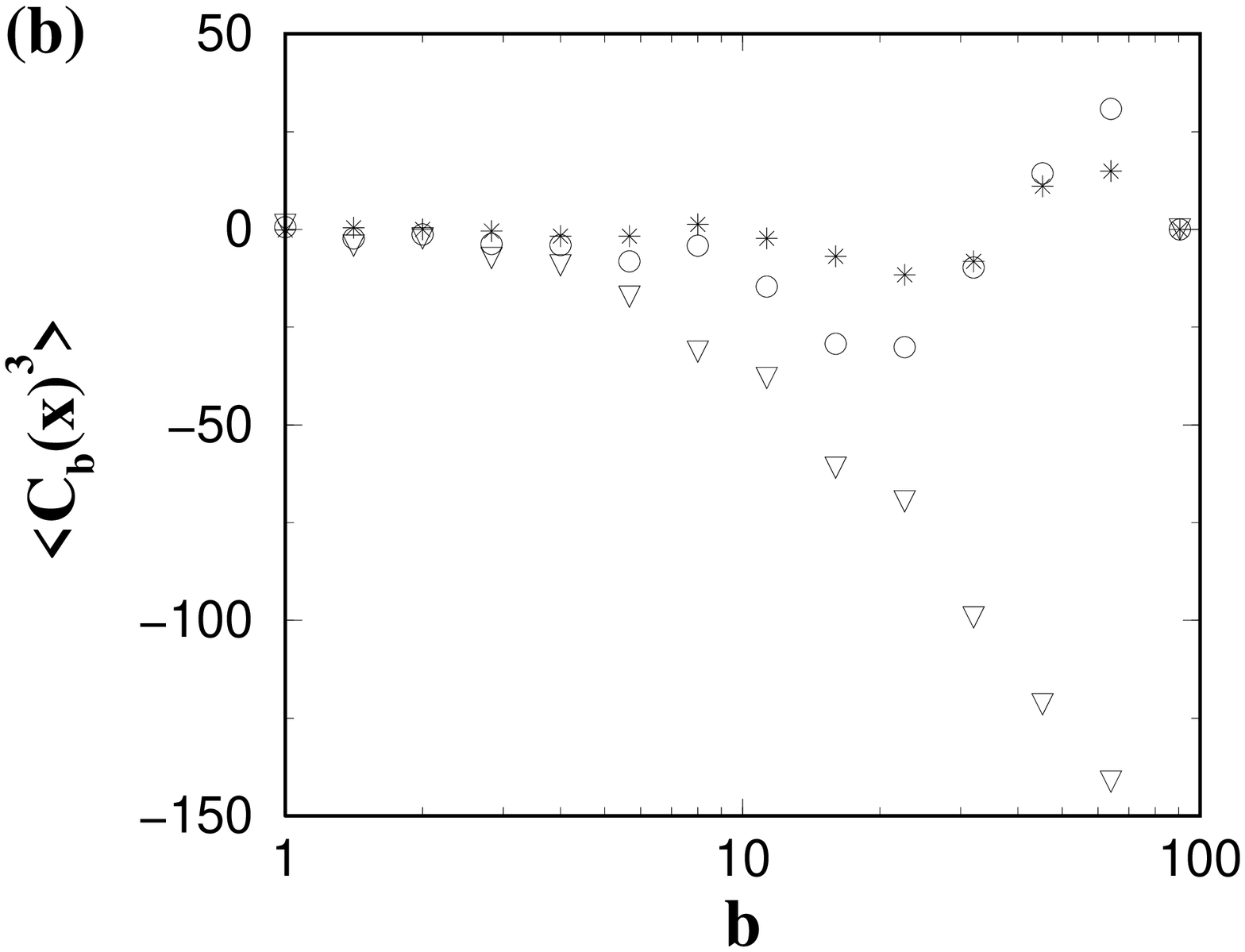}
   \caption{
Second moment (a) and third moment (b) 
of the scale-dependent curvature $C_b({\bf x})$, 
for surfaces from the single-step model with $p_+=0.1$ ($\bigtriangleup$), 
$p_+=0.3$ ($\circ$), and $p_+=0.5$ ($\ast$). 
In (b), the $p_+=0.3$ and $0.5$ data are consistent with
$\langle C_b^3 \rangle =0$, while the
$p_+=0.1$ data
show a strong (and non-Gaussian) breaking of up/down symmetry.}
   \label{Fig:SSM_C}
  \end{minipage}
\end{figure}

\begin{figure}
  \begin{minipage}{8.66cm}
   \epsfxsize=8.66cm \epsfbox{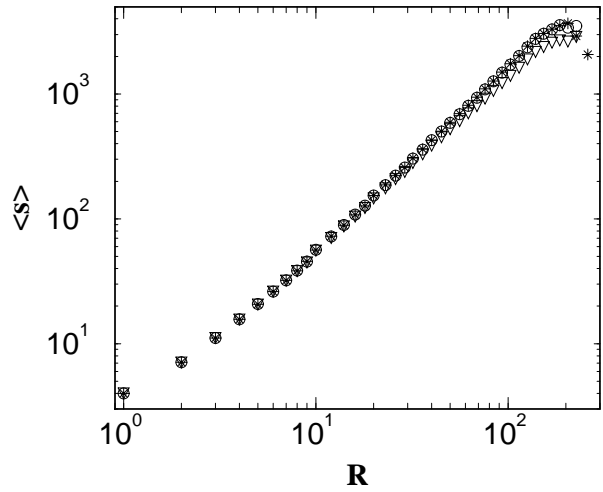}
   \caption{
Average loop length ``vs'' the loop radius for $p_+=0.1$ ($\bigtriangleup$), 
$p_+=0.3$ ($\circ$), and $p_+=0.5$ ($\ast$), in the single-step model.
Note that the  $p_+=0.3$ and
$p_+=0.5$ data are almost indistinguishable.  
           }
   \label{Fig:SSM_SR}
  \end{minipage}
\end{figure}

\begin{figure}
  \begin{minipage}{8.66cm}
   \epsfxsize=8.66cm \epsfbox{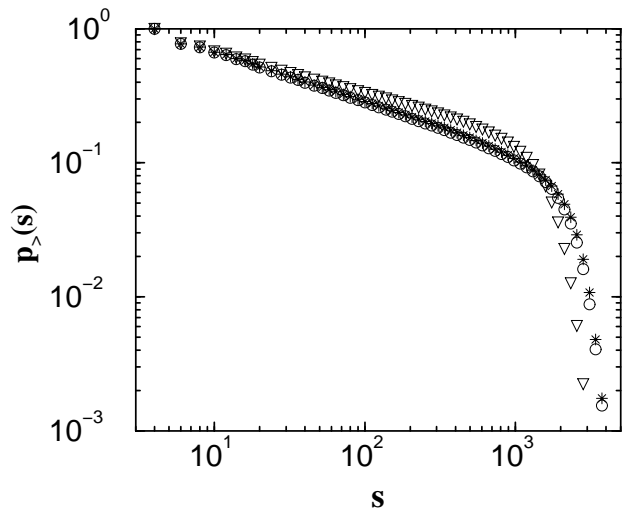}
   \caption{
Normalized loop size distribution 
for $p_+=0.1$ ($\bigtriangleup$), 
$p_+=0.3$ ($\circ$), and $p_+=0.5$ ($\ast$), in the 
single step model. Again, the  $p_+=0.3$ and
$p_+=0.5$ plots are almost indistinguishable.  
           }
   \label{Fig:SSM_PS}
  \end{minipage}
\end{figure}

\begin{figure}
  \begin{minipage}{8.66cm}
   \epsfxsize=8.66cm \epsfbox{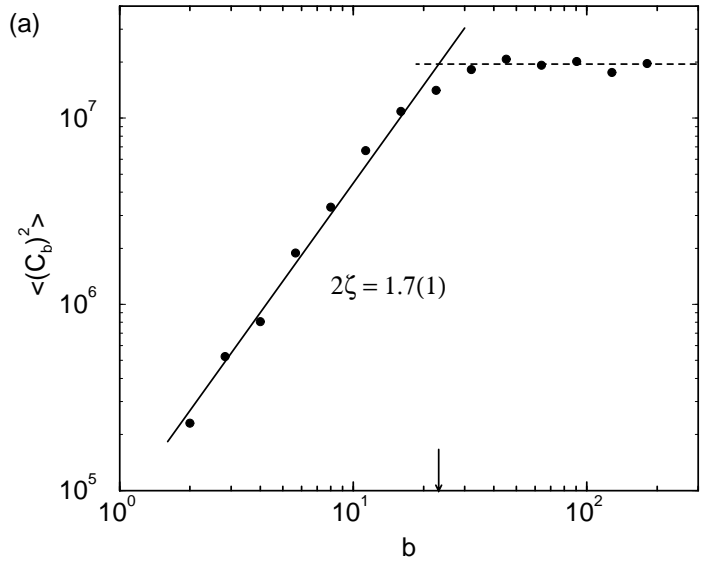}
   \epsfxsize=8.66cm \epsfbox{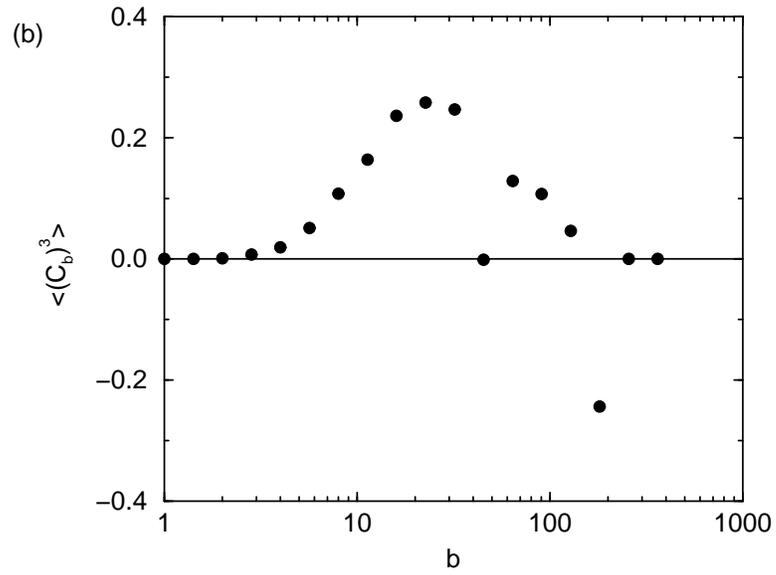}
   \caption{
Second moment (a) and third moment (b) of
the scale-dependent curvature, as evaluated for
a 702 nm thick Ag film, grown on quartz, 
from the STM data of Palasantzas and Krim (Ref.~\protect\onlinecite{krim}).
           }
   \label{Fig:Cb_Krim}
   \label{fig_krim_C}
  \end{minipage}
\end{figure}

\begin{figure}
  \begin{minipage}{8.66cm}
   \epsfxsize=8.66cm \epsfbox{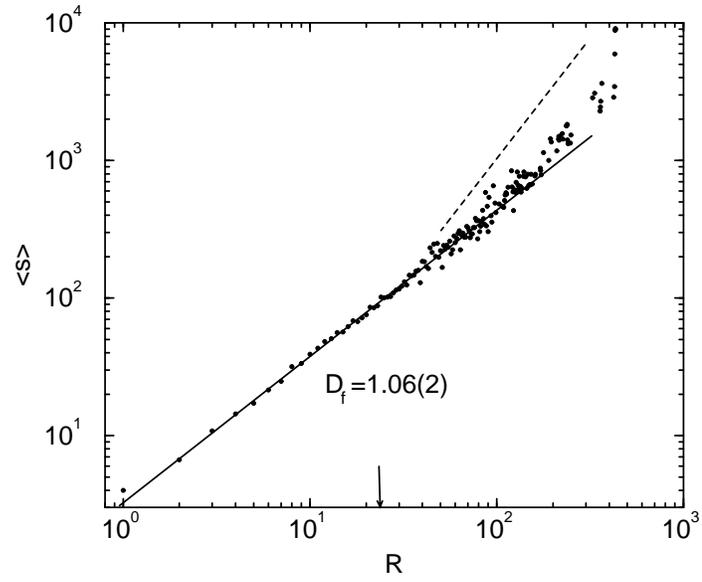}
   \caption{
Mean contour length $\langle s \rangle$ as a function of
radius $R$, for the Ag film of Ref.~\protect\onlinecite{krim}.
Here 1000 contour loops were collected from the STM data 
of Ref.~\protect\onlinecite{krim}.
The solid line is the least-squares  best 
fit for radii $2<R<125$; its slope is the estimated fractal dimension $D_f$.
The slope of the dashed line is equal to the hull dimension of 
critical percolation clusters. }
\label{Fig:sR_Krim}
\label{fig:PKdata}
  \end{minipage}
\end{figure}

\begin{figure}
  \begin{minipage}{8.66cm}
   \epsfxsize=8.66cm \epsfbox{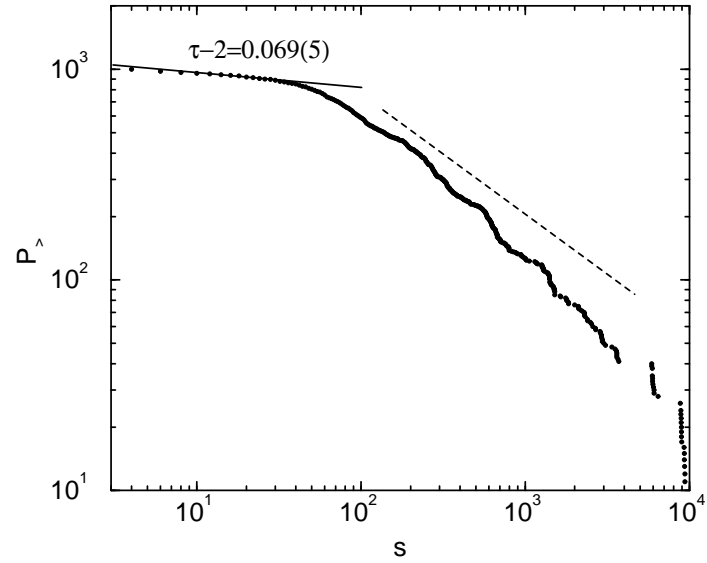}
   \caption{
Cumulative distribution of contour loop lengths from 
STM data, for the Ag film of Ref.~\protect\onlinecite{krim}.
The solid line is the result of a linear fit to the data in 
the affine-scaling regime. The slope of the dashed line corresponds
to the exponent $\tau-2$ in the percolation regime. }
   \label{Fig:PS_Krim}
  \end{minipage}
\end{figure}

\begin{figure}
\begin{minipage}{8.66cm}
\epsfxsize=8.66cm \epsfbox{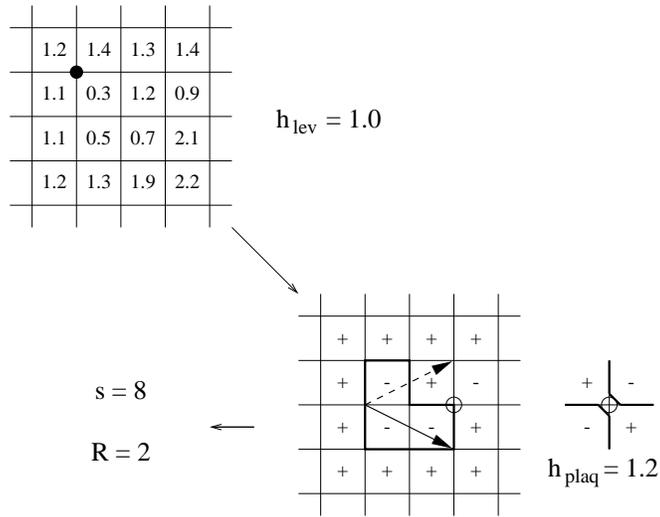}
\caption{Construction of contour loops of a random surface
on a lattice. Heights $h({\bf r})$ are indicated by numbers in the cells; 
$h_{\rm lev}$ is the height of the level set through the chosen point
(filled circle) while $h_{\rm plaq}$ is the height of the ``saddle-point''
(unfilled circle).   
Our definition of the diameter $R$ and loop length $s$ is indicated.
The solid arrow connects points on the same loop, and thus 
contributes to the loop correlation function $G({\bf r})$; the
dashed arrow does {\it not} contribute: it connects points of 
the same level set, but they are on disconnected loops.}
\label{fig_walks}
\end{minipage}
\end{figure}

\end{centering}


\begin{references}
\bibitem{FN_water} For lattice models that map to logarithmically rough 
Gaussian surfaces, the probability that 
two loops with radii larger than $R$ have a distance of closest approach  
smaller than  $r$ vanishes as a power of $r/R$
(Ref.~\onlinecite{saleur}).
\bibitem{FN-nonlocalSSM}
This makes the algorithm nonlocal, since the rate at which
deposition occurs on an allowed site depends on the number of such sites
in the rest of the system.
\bibitem {racz}
The ensemble distribution functions (for many independent samples) 
of the system-wide height variance is another characterization of
roughness; see Ref.~\onlinecite{plish}.
\end{references}
\end{document}